\documentclass[journal,onecolumn]{IEEEtran}
\usepackage{amsmath}
\usepackage{amssymb}
\usepackage{graphicx}
\usepackage{enumerate}
\usepackage{comment}
\usepackage{mathtools}
\usepackage{scrextend} 

\newtheorem{thm}{Theorem}
\newtheorem{lemma}[thm]{Lemma}

\newtheorem{proposition}[thm]{Proposition}

\def\QEDclosed{\mbox{\rule[0pt]{1.3ex}{1.3ex}}} 

\begin{document}


\title{Learning to detect an oddball target}

\author{\IEEEauthorblockN{Nidhin Koshy Vaidhiyan} and \IEEEauthorblockN{Rajesh Sundaresan} \thanks{This work was supported by the Indo-French Centre for the Promotion of Advanced Research under grant No. 5100-ITA.}}

\maketitle

\begin{abstract}
 We consider the problem of detecting an odd process among a group of Poisson point processes, all having the same rate except the odd process. The actual rates of the odd and non-odd processes are unknown to the decision maker. We consider a time-slotted sequential detection scenario where, at the beginning of each slot, the decision maker can choose which process to observe during that time slot. We are interested in policies that satisfy a given constraint on the probability of false detection. We propose a generalised likelihood ratio based sequential policy which, via suitable thresholding, can be made to satisfy the given constraint on the probability of false detection. Further, we show that the proposed policy is asymptotically optimal in terms of the conditional expected stopping time among all policies that satisfy the constraint on the probability of false detection. The asymptotic is as the probability of false detection is driven to zero. 
 
 We apply our results to a particular visual search experiment studied recently by neuroscientists. Our model suggests a neuronal dissimilarity index for the visual search task.   The neuronal dissimilarity index, when applied to visual search data from the particular experiment, correlates strongly with the behavioural data. However, the new dissimilarity index performs worse than some previously proposed neuronal dissimilarity indices. We explain why this may be attributed to the experiment conditons.
\end{abstract}

\section{Introduction}
 Consider $K$ homogeneous Poisson point processes. All processes except one, which we call the ``odd'' process, have the same rate. The actual rates of the odd process and the non-odd processes are unknown. The objective is to detect the odd (or anomalous or outlier) process as quickly as possible, but subject to constraints on the probability of false detection. For simplicity, we assume that time is divided into slots of fixed duration $T$. During a particular time slot, the decision maker can choose exactly one among the $K$ processes for observation. This choice is made only at slot beginnings. 
 
 We cast the above problem into one of sequential detection with control \cite{ref:195909AMS_Che}, but with unknown underlying distributions. The structural constraints in the problem, that exactly one among the $K$ processes has a distribution different from the others, opens up an opportunity to {\it learn} about the underlying distributions from the observations, and yet, learn just about enough to make a reliable decision.
 
  We adapt the sample complexity result of Kaufmann et al. \cite{ref:kaufmann2014complexity}, originally developed for the best arm identification problem, to our setting and obtain a lower bound on the conditional expected stopping time for any policy that satisfies the constraint on the probability of false detection. The lower bound suggests that the conditional expected stopping time is asymptotically proportional to the negative of the logarithm of the probability of false detection. The proportionality constant is obtained as the solution to a max-min optimisation problem of relative entropies between the true system state (index of the odd process, its rate, and the rate of the non-odd processes) and other alternatives.  The optimisation problem for the lower bound also suggests the nature of an asymptotically optimal strategy. 
  
 
 The usual methodology employed in problems with lack of exact knowledge of the underlying distributions is to use tests that are based on generalised likelihood ratios (GLR tests or GLRT). We work with a modification of the GLRT. Unlike the usual GLRT statistic, we replace the maximum likelihood function in the numerator of the statistic by an average likelihood function, the average computed with respect to an artificial prior on  the odd and non-odd rates. For the Poisson model, we employ a gamma distribution on the rates of the odd and non-odd processes as the prior, with the shape and rate parameter set to one. In fact, any prior density having full support would suffice. The specific gamma prior allows easier characterisation of the averaged likelihood function.  The averaging prevents over estimation of the likelihood ratio function, and at the same time ensures that, asymptotically, the averaged version is not too far away from the true likelihood function.  The modification allows us to design a time invariant and simple threshold policy that satisfies the probability of false detection constraint. We show that the sampling strategy of the proposed policy (which of the $K$ processes to observe at the beginning of each slot) converges to the sampling strategy suggested by the lower bound, where the convergence is as the number of slots observed tends to infinity. We show that, asymptotically, the conditional expected stopping time of the proposed policy scales as $-\log (P_{e})/ D^{*}$, where $P_{e}$ is the constraint on the probability of false detection  and $D^{*}$, a relative entropy based constant, is the optimal scaling factor as suggested by the lower bound.
 
 The motivation to study this problem comes from a visual search problem studied by Sripati and Olson \cite{ref:201001JNS_SriOls}, where a subject has to detect an odd image among a sea of distractor images ``as quickly as possible without guessing''  \cite{ref:201001JNS_SriOls}. We model the visual search task as an oddball detection problem, as above, and propose $D^{*}$ as a neuronal dissimilarity index for such visual search tasks. We compare the performance of the proposed dissimilarity index with other dissimilarity indices proposed earlier by Vaidhiyan et al. in  \cite{ref:VaiArunSund_Journal1_Arxiv_201506}. In that paper, it was assumed that the odd and the non-odd rates were known. Our proposed dissimilarity index of this paper correlates strongly with some behavioural data of \cite{ref:201001JNS_SriOls}. However, the proposed dissimilarity index performs slightly worse than the neuronal dissimilarity index proposed by Vaidhiyan et al. in \cite{ref:VaiArunSund_Journal1_Arxiv_201506}. Nevertheless, we present the comparisons on the existing experimental data. 
 
 \subsection{Prior Work}
 Sequential hypothesis testing with control, assuming knowledge of the underlying distributions of the observations under different hypotheses, was first studied by  Chernoff \cite{ref:195909AMS_Che}. Such problems are also known as Active Sequential Hypothesis Testing Problems (ASHT) \cite{ref:201006ISIT_NagJav, ref:201312TAS_NagJav}. Chernoff \cite{ref:195909AMS_Che} studied ASHT in the context of designing optimal experiments. His performance criterion was the total cost of sampling, which is proportional to delay, plus a penalty for false detection. Chernoff proposed a policy, the so-called {\it Procedure A},  and showed its asymptotic optimality as the cost of sampling went to zero. {\it Procedure A} maintains a posterior distribution on the set of hypotheses and, at each instant, selects actions according to the hypothesis with the highest posterior probability. In a series of works, Naghshvar and Javidi \cite{ref:201006ISIT_NagJav, ref:201312TAS_NagJav, ref:201010Alt_NagJav, ref:201108ISIT_NagJav, ref:201310JSP_NagJav} studied ASHT from a Bayesian cost minimisation perspective. Nitinawarat et al. \cite{ref:201310ITAC_NitinVeeravalli, nitinawarat2015controlled} studied ASHT from the perspective of  minimising the conditional expected cost (generally stopping delay), subject to constraints on the probability of false detection. All the above works assumed knowledge of the underlying distributions under different hypotheses. 

 Li et al. \cite{ref:LiNitinVeer_Universal_Outlier_Hypothesis_Testing_TIT201407} studied fixed sample size outlier detection under unknown typical and outlier distributions, but in a finite observation space setting. They assumed simultaneous observability of all processes at each observation instance. They proposed a modified GLRT which was shown to have, asymptotically, the same error exponent as that of an optimal algorithm with knowledge of the underlying distributions. The asymptotics was as the number of processes available for observation tended to infinity. They termed such algorithms {\it asymptotically exponentially consistent}. Further, they extended their study to the setting where there are more than one outlier processes. They extended their algorithm and showed that it is asymptotically exponentially consistent in the new setting. Li et al. \cite{ref:LiNitinVeer_Universal_Sequential_Outlier_Hypothesis_Testing_ISIT2014} studied sequential versions of  \cite{ref:LiNitinVeer_Universal_Outlier_Hypothesis_Testing_TIT201407} and showed that another modified GLRT that  keeps sampling until the test statistic crosses a threshold is universally consistent as the threshold is increased to infinity. In both these works, unlike in the ASHT setting and unlike our setting, at each observation instance, observations from all the processes were available to the decision maker. 
Nitinawarat and Veeravalli \cite{ref:NitinVeer_Universal_Scheme_for_Optimal_Search_and_Stop_2014arXiv1412.4870N} studied an outlier detection problem in a setting similar to the one being considered in this paper, where at each observation instance, the decision maker is allowed to observe only one of the processes. But different from our setting, they assume knowledge of the typical (or non-odd) distribution. They proposed an algorithm that was shown to have vanishing probability of false detection as the threshold is increased to infinity. Further, the proposed algorithm was shown to have, asymptotically, the same error exponent as that of an optimal policy with knowledge of the atypical (odd) distribution. Recently, Cohen and Zhao \cite{ref:CohenZhao_Asymptotically_Optimal_Anomaly_Detection_via_Sequential_Testing_TSP_Jun2015} studied a problem similar to ours, but restricted their study to the setting when the atypical (odd) and typical (non-odd) distributions belonged to disjoint parameter sets. Consequently, in their setting, the optimal action at each decision instance is to observe the process that has the generalised maximum likelihood with respect to the set of atypical (odd) parameters. Their proposed policy also had a threshold based stopping criterion. They showed that their policy has the same asymptotic scaling for the conditional expected stopping time as for an optimal policy with knowledge of the distributions. 
 
 A related problem, studied extensively by the machine learning community, is the problem of identification of the {\it best arm} for multi-armed bandits. Kaufmann et. al \cite{ref:kaufmann2014complexity} studied the sample complexity of the best arm identification problem. Our problem of anomaly detection can be cast as an odd-arm identification problem. The structures in the problems that need to be exploited are different. 
 
 \subsection{Our Contribution}
 
 Our asymptotically optimal algorithm differs from those in prior works in the following aspects:
 \begin{itemize}
  \item Unlike the works on ASHT \cite{ref:195909AMS_Che, ref:201006ISIT_NagJav, ref:201312TAS_NagJav, ref:201010Alt_NagJav, ref:201108ISIT_NagJav, ref:201310JSP_NagJav, ref:201310ITAC_NitinVeeravalli, nitinawarat2015controlled}, we assume no knowledge of the underlying distribution under different hypotheses. However, our proposed algorithm is an adaptation of Chernoff's {\it Procedure A}  to our setting.
  \item For a given probability of false detection constraint, we propose a policy with a new modification of GLRT and a fixed threshold such that it satisfies the constraint.
  \item Unlike the works of Li et al. \cite{ref:LiNitinVeer_Universal_Outlier_Hypothesis_Testing_TIT201407}, \cite{ref:LiNitinVeer_Universal_Sequential_Outlier_Hypothesis_Testing_ISIT2014}, our observations are limited by the chosen actions. There is then a clear exploration versus exploitation tradeoff.
  \item Unlike the work of Nitinawarat and Veeravalli \cite{ref:NitinVeer_Universal_Scheme_for_Optimal_Search_and_Stop_2014arXiv1412.4870N}, we do not assume knowledge of the atypical (odd) distribution, nor do we assume the typical (non-odd) distribution.
  \item Unlike the work of  Cohen and Zhao \cite{ref:CohenZhao_Asymptotically_Optimal_Anomaly_Detection_via_Sequential_Testing_TSP_Jun2015}, we do not assume that the atypical and typical distributions belong to disjoint sets.
  \item We specifically consider the setting of Poisson point processes mainly because of our desire to explain the experimental observations of Sripati and Olson \cite{ref:201001JNS_SriOls} on neuronal data which are modelled as Poisson point processes in \cite{ref:VaiArunSund_Journal1_Arxiv_201506}. Nevertheless, we believe that the same ideas may carry forward to other class of distributions, especially exponential families. 
 \end{itemize}

 \subsection{Organisation}
 In Section \ref{sec: Modelling and Notation}, we develop the required notation and describe the model. In Section \ref{sec: The converse - lower bound}, we provide a lower bound on the conditional expected stopping time for any policy that satisfies the probability of false detection constraint. The nature of the lower bound suggests a candidate asymptotically optimal policy. In the same section, we  make some observations on some structural properties of the suggested policy. In Section \ref{sec: upper bound}, we formally propose the policy and show that it is asymptotically optimal. In Section \ref{sec: Application to Visual Search}, we apply the theory to visual search. We show that the proposed neuronal dissimilarity index is strongly correlated with the behavioural data. In Section \ref{sec: Conclusion}, we make some concluding remarks and discuss possible extensions. Most proofs are relegated to appendices \ref{proof: Proof of lambda bounded away from 0 and 1} and \ref{appendix: Properties of policy pi MGLRT}.

 \section{Model}
 \label{sec: Modelling and Notation}
 In this section we develop the required notation and describe the model.
 
 Let $K \ge 3$ denote the number of Poisson point processes under consideration. Conditioned on the rates, the processes are assumed to be independent of each other. Let $H, 1 \le H \le K$, denote the index of the odd process. Let $R_{1} > 0$ denote the unknown rate of the odd process, and let $R_{2} > 0$ denote the unknown rate of the non-odd processes. We assume $R_{1} \ne R_{2}$. Let the triplet $\Psi = (H, R_{1}, R_{2})$ denote the configuration of the processes, where the first component represents the index of the odd process, while the second and third components represent the odd and non-odd rates respectively. Let $T$ denote the time duration of a time slot. Without loss of generality we can assume $T = 1$, the analysis holds for general $T$ with an appropriate scaling of the rates. The analysis can be done in continuous time as well, but we shall take the simpler slotted time approach.
 
 Given the Poisson process assumption, a sufficient statistic for the observed process during a time slot is the number of jumps observed in that time slot. Let $A_{n} \in \{1, 2, \ldots, K\}$ denote the index of the process chosen for observation in time slot $n$, and let $X_{n} \in  \mathbb{Z}_{+}$ denotes the number of jumps observed in the process during time slot $n$. Let $(X_{n})_{n \ge 1}$ and $(A_{n})_{n \ge 1}$ denote the observation process and the control process respectively. We write $X^{n}$ for $(X_{1}, X_{2}, \ldots, X_{n})$ and $A^{n}$ for $(A_{1}, A_{2}, \ldots, A_{n})$. We also write $\mathcal{P}(K)$ for the set of probability distributions on $\{1, 2, \ldots, K\}$.

 A policy $\pi$ is a sequence of action plans that at time $n$ looks at the history $X^{n-1}, A^{n-1}$ and prescribes a composite action that is either $(stop, \delta)$ or $(continue, \lambda)$ as explained next. If the composite action is $(stop, \delta)$, then the detector stops taking further samples (or retires) and indicates $\delta$ as its decision on the hypotheses; $\delta \in \{1, 2, \dots, K\}$. If the composite action is $(continue, \lambda)$, the detector picks the next action $A_{n}$ according to the distribution $\lambda \in \mathcal{P}(K)$. The stopping time is defined as $$\tau := \inf \{n \ge 1 : A_{n} = (stop, \cdot)\}.$$

 Consider a policy $\pi$. Conditioned on action $A_{n}$, the true hypothesis $H$, and the odd and non-odd rates $R_{1}$ and $R_{2}$, we assume that the observation $X_{n}$ is independent of previous actions $A^{n-1}$, previous observations $X^{n-1}$, and the policy. The conditional distribution of $X_{n}$, given the current action $A_{n}$, the configuration $\Psi = (H,R_{1},R_{2})$, the history $X^{n-1}, A^{n-1}$, and  the Poisson assumption, is given by
\begin{align}
 P(X_{n} = l \vert \Psi = (H,R_{1}, R_{2}), A_{n},X^{n-1}, A^{n-1}) & =  P(X_{n} = l \vert \Psi = (H,R_{1}, R_{2}), A_{n})\\
 & =
 \begin{cases}
  \frac{R_{1}^{l} e^{-R_{1}}}{l!} &\text{ if } A_{n} = H \\
  \frac{R_{2}^{l} e^{-R_{2}}}{l!} &\text{ if } A_{n} \ne H,
 \end{cases}
\end{align}
where $l \in \mathbb{Z}_{+}$.

Let $E^{\pi}$ denote the conditional expectation and let $P^{\pi}$ denote the conditional probability measure, given $\Psi$, under the policy $\pi$. Given an error tolerance vector $\alpha = (\alpha_{1}, \alpha_{2}, \ldots, \alpha_{n})$, with $0 < \alpha_{i}<1$, let $\Pi(\alpha)$ be the set of desirable policies defined as
\begin{align}
 \Pi(\alpha) := \{\pi : P^{\pi}(\delta \ne i | \Psi = (H, R_{1}, R_{2}), H = i) \le \alpha_{i}, \text{ for all } i \text{ and for all } \Psi \text{ such that } R_{1} \ne R_{2}\}.
\end{align}
Let $\Vert \alpha \Vert$ denote $\max_{i}{\alpha_{i}}$.

For ease of notation, we drop the superscript $\pi$ while writing $E^{\pi}$, $P^{\pi}$, and other variables, but their dependence on the underlying policy should be kept in mind, and the policy under consideration will be clear from the context.

\section{The converse - Lower bound}
\label{sec: The converse - lower bound}
In this section we develop a lower bound on the conditional expected stopping time for any policy that belongs to $\Pi(\alpha)$. We show that, as $\Vert \alpha \Vert \rightarrow 0 $, the lower bound scales as $- \log(\Vert \alpha \Vert) / D^{*}$. We also characterise $D^{*}$ in detail in this section. 

The following proposition gives a lower bound on the conditional expectation of the stopping time for all policies belonging to $\Pi(\alpha)$. The proof may be seen as an application of the data processing inequality (\cite[p. 16]{ref:kullback1959information}, \cite{ref:AliSilvey_1966}) for relative entropy. 
\vspace*{0.1 in}
\begin{proposition}
\label{prop: Lower bound}
 Fix $\alpha$, with $ 0<\alpha_{i}<1$ for each $i$. Let $\Psi = (i, R_{1}, R_{2})$ be the true configuration. For any $\pi \in \Pi(\alpha)$, we have
 \begin{align}
  \label{eqn: prop lower bound}
  E^{\pi}\left[\tau \vert \Psi \right] \ge \frac{d_{b}(\Vert \alpha \Vert, 1-\Vert \alpha \Vert)}{D^{*}(i, R_{1}, R_{2})},
 \end{align}
 where $d_{b}(\Vert \alpha \Vert, 1-\Vert \alpha \Vert)$ is the binary relative entropy function defined as $$d_{b}(x, 1-x) := x \log(x/(1-x))+(1-x) \log((1-x)/x),$$ and $D^{*}(i, R_{1}, R_{2})$ is defined as
 
 \begin{align}
 \label{eqn:D^{*} original form}
  D^{*}(i, R_{1},R_{2}) := \max_{\lambda \in \mathcal{P}(K)} \min_{R_{1}' > 0, R_{2}' > 0, j \ne i} \left[ \lambda(i) D(R_{1} \Vert R_{2}')+ \lambda(j) D(R_{2} \Vert R_{1}') + (1-\lambda(i)-\lambda(j)) D(R_{2} \Vert R_{2}')\right],
 \end{align}
where $D(x \Vert y) := x \log(x/y) -x + y$ is the KL-divergence or relative entropy between two Poisson random variables with means $x$ and $y$.
\end{proposition}
\vspace*{0.1 in}

Let $\lambda^{*}(i, R_{1}, R_{2})$ denote the $\lambda \in \mathcal{P}(K)$ that maximises (\ref{eqn:D^{*} original form}), i.e.,
\begin{align}
\label{eqn: lambda star original form}
 \lambda^{*}(i, R_{1}, R_{2}) = \arg \max_{\lambda \in \mathcal{P}(K)} \min_{R_{1}', R_{2}', j \ne i} \left[ \lambda(i) D(R_{1} \Vert R_{2}')+ \lambda(j) D(R_{2} \Vert R_{1}') + (1-\lambda(i)-\lambda(j)) D(R_{2} \Vert R_{2}')\right].
 \end{align}
 
 We can interpret $D^{*}(i, R_{1}, R_{2})$ as the minimum among relative entropy rates between the true configuration $\Psi = (i, R_{1}, R_{2})$ and all other possible alternate configurations $\Psi' = (j,R_{1}', R_{2}')$, with  $j \ne i$, but maximised over all policies (action strategies) that pick actions in an independent and identically distributed (i.i.d.) manner. It can also be interpreted as the max-min-drift of the log likelihood ratio process between the true configuration and other error configurations, the minimum being over all possible error configurations, and the maximum being over all i.i.d. policies. $D^{*}(i, R_{1}, R_{2})$ is the key information quantity in this paper. Since $d_{b}(\Vert \alpha \Vert, 1-\Vert \alpha \Vert) / \log(\Vert \alpha \Vert) \rightarrow 1$ as $\Vert \alpha \Vert \rightarrow 0$, Proposition \ref{prop: Lower bound} shows that the conditional expected stopping time of the optimal policy scales at least as $-\log(\Vert \alpha \Vert)/D^{*}(i, R_{1}, R_{2})$ as the probability of false detection constraint $\Vert \alpha \Vert \rightarrow 0$. In Section \ref{sec: upper bound} we will describe a policy that is upper bounded by, and therefore achieves, a similar scaling, though only asymptotically as $\Vert \alpha \Vert \rightarrow 0$.

\vspace*{0.1 in}
\begin{IEEEproof}[Proof of Proposition \ref{prop: Lower bound}]
Assume $E^{\pi}\left[\tau \vert \Psi \right]$ is finite, for otherwise (\ref{eqn: prop lower bound}) is trivially true. We apply the sample complexity result of Kaufmann et al. \cite[Lemma 1]{ref:kaufmann2014complexity} to our setting. Let $N_{j}(\tau) = \sum_{k =1}^{\tau} 1_{\{A_{k} = j\}}$ denote the number of samples from process $j$ observed till the stopping time $\tau$. Clearly,  $\tau = \sum_{j =1}^{K} N_{j}(\tau)$.  Kaufmann et al. \cite[Lemma 1]{ref:kaufmann2014complexity} showed that, for any $\pi \in \Pi(\alpha)$, conditioned on the true configuration $\Psi = (i, R_{1}, R_{2})$, and for any alternate configuration $\Psi' = (j, R_{1}', R_{2}') , j \ne i$, the conditional expected sample sizes satisfy
 \begin{align}
 \label{eqn:Kaufmann Lower bound}
  E^{\pi}\left[N_{i}(\tau) \vert \Psi \right] D(R_{1} \Vert R_{2}')+ E^{\pi}\left[N_{j}(\tau) \vert \Psi \right] D(R_{2} \Vert R_{1}') + \left(\sum_{k \ne i, k \ne j} E_{i}^{\pi}\left[N_{k}(\tau) \vert \Psi \right]\right) D(R_{2} \Vert R_{2}') \ge d_{b}(\Vert \alpha \Vert, 1-\Vert \alpha \Vert).
 \end{align}
 Multiplying and then dividing  the left-hand side by $E^{\pi}\left[\tau \vert \Psi \right]$, we get
\begin{align}
\nonumber d_{b}&(\Vert \alpha \Vert, 1-\Vert \alpha \Vert)\\
\label{eqn: Lower bound 1}& \le  E^{\pi}\left[\tau \vert \Psi \right] \left[\frac{E^{\pi}\left[N_{i}(\tau) \vert \Psi \right]}{E^{\pi}\left[\tau \vert \Psi \right]} D(R_{1} \Vert R_{2}')+ \frac{E^{\pi}\left[N_{j}(\tau) \vert \Psi \right]}{E^{\pi}\left[\tau \vert \Psi \right]} D(R_{2} \Vert R_{1}') + \left(1-\frac{E^{\pi}\left[N_{i}(\tau) \vert \Psi \right]+E^{\pi}\left[N_{j}(\tau) \vert \Psi \right]}{E^{\pi}\left[\tau \vert \Psi \right]}\right) D(R_{2} \Vert R_{2}')\right].
\end{align}
Since (\ref{eqn: Lower bound 1}) holds for any $R_{1}', R_{2}'$ and $j \ne i$, and since $E^{\pi}\left[\tau \vert \Psi \right]$ does not depend on $R_{1}', R_{2}'$ and $j \ne i$, we can choose the tightest bound and get

\begin{align}
\nonumber d_{b}&(\Vert \alpha \Vert, 1-\Vert \alpha \Vert)\\
\nonumber & \le E^{\pi}\left[\tau \vert \Psi \right] \min_{R_{1}', R_{2}', j \ne i}  \left[\frac{E^{\pi}\left[N_{i}(\tau) \vert \Psi \right]}{E^{\pi}\left[\tau \vert \Psi \right]} D(R_{1} \Vert R_{2}')+ \frac{E^{\pi}\left[N_{j}(\tau) \vert \Psi \right]}{E^{\pi}\left[\tau \vert \Psi \right]} D(R_{2} \Vert R_{1}') \right. \\ 
\label{eqn: Lower bound 2} & \hspace*{6 cm} \left. + \left(1-\frac{E^{\pi}\left[N_{i}(\tau) \vert \Psi \right]+E^{\pi}\left[N_{j}(\tau) \vert \Psi \right]}{E^{\pi}\left[\tau \vert \Psi \right]}\right) D(R_{2} \Vert R_{2}')\right]\\
\label{eqn: Lower bound 3} & \le E^{\pi}\left[\tau \vert \Psi \right] \max_{\lambda \in \mathcal{P}(K)} \min_{R_{1}', R_{2}', j \ne i}  \left[\lambda(i) D(R_{1} \Vert R_{2}')+ \lambda(j) D(R_{2} \Vert R_{1}') + (1-\lambda(i)-\lambda(j)) D(R_{2} \Vert R_{2}')\right].
\end{align}
The last inequality follows because maximisation over all $\lambda \in \mathcal{P}(K)$ only increases the right-hand side. This completes the proof.
\end{IEEEproof}
\vspace*{0.1 in}

 We now describe some simplifications for $D^{*}(i, R_{1}, R_{2})$ and $\lambda^{*}(i, R_{1}, R_{2})$. We show that the $K$-dimensional optimisation in (\ref{eqn:D^{*} original form}) can be reduced to a one-dimensional optimisation. 
 \vspace*{0.1 in}
 \begin{proposition}
  Consider $K$ Poisson point processes with configuration $\Psi = (i, R_{1}, R_{2})$. The quantity $D^{*}(i, R_{1}, R_{2})$ of (\ref{eqn:D^{*} original form})  can be equivalently expressed as
  \begin{align}
  \label{eqn: D star simpler form}
   D^{*}(i, R_{1}, R_{2}) = \max_{0 \le \lambda(i) \le 1} \left[\lambda(i) D(R_{1} \Vert \tilde{R}) + (1-\lambda(i)) \frac{(K-2)}{(K-1)} D(R_{2} \Vert \tilde{R})\right],
  \end{align}
where 
\begin{align}
\label{eqn: R tilde}
 \tilde{R} = \left( \lambda(i) R_{1} + (1-\lambda(i))\frac{(K-2)}{(K-1)} R_{2} \right) / \left( \lambda(i) + (1-\lambda(i))\frac{(K-2)}{(K-1)} \right).
\end{align}
Also, $\lambda^{*}(i, R_{1}, R_{2})$ is of the form
  \begin{align}
   \lambda^{*}(i, R_{1}, R_{2})(j) &=
   \begin{cases}
    \lambda^{*}(i,R_{1},R_{2})(i) &\text{ if } j = i\\
    (1-\lambda^{*}(i,R_{1},R_{2})(i))/(K-1) &\text{ if } j \ne i.
   \end{cases}
  \end{align}
   \end{proposition}

 \begin{IEEEproof}
 Consider (\ref{eqn:D^{*} original form}). Observe that $R_{1}'$ appears only in the middle term on the right-hand side. This is minimised when $R_{1}' = R_{2}$ and the minimum value is zero. We therefore  have
  \begin{align}
  \label{eqn:D* simplification 2} D^{*}(i, R_{1},R_{2})  & = \max_{\lambda \in \mathcal{P}(K)} \min_{R_{2}', j \ne 1} \left[ \lambda(i) D(R_{1} \Vert R_{2}')+ (1-\lambda(i)-\lambda(j)) D(R_{2} \Vert R_{2}')\right]\\
  \label{eqn:D* simplification 3} & = \max_{ 0 \le \lambda(i) \le 1} \min_{R_{2}'} \left[\lambda(i) D(R_{1} \Vert R_{2}')+ (1-\lambda(i)) \frac{(K-2)}{(K-1)} D(R_{2} \Vert R_{2}')\right]. 
 \end{align} 
 Equation  (\ref{eqn:D* simplification 3}) follows from the fact that the $\lambda$ that maximises (\ref{eqn:D* simplification 2}) will have equal mass on all locations other than $i$, i.e., the maximiser $\lambda^{*}$ will satisfy $\lambda^{*}(j) = (1-\lambda^{*}(i))/(K-1), ~ \text{for all } j \ne i$.
  
 For a fixed $\lambda(i)$, to find the $R_{2}'$ that minimises the term within brackets in (\ref{eqn:D* simplification 3}) which is a strictly convex function of $R_{2}'$, we take its derivative with respect to $R_{2}'$ and equate it to zero. We then see that the minimising $R_{2}'$ satisfies the equation
\begin{align}
\label{eqn: D derivative equal to zero to obtain R}
 \lambda(i) D'(R_{1} \Vert R_{2}')+ (1-\lambda(i)) \frac{(K-2)}{(K-1)} D'(R_{2} \Vert R_{2}') = 0,
\end{align}
where $D'(x \Vert y)$ is the derivative of $D(x \Vert y)$ with respect to the second argument $y$, which turns out to be $1-x/y$.
The $R_{2}'$ thus obtained is
\begin{align}
 R_{2}' = \left( \lambda(i) R_{1} + (1-\lambda(i))\frac{(K-2)}{(K-1)} R_{2} \right) / \left( \lambda(i) + (1-\lambda(i))\frac{(K-2)}{(K-1)} \right).
\end{align}
This completes the proof.
\end{IEEEproof}
\vspace*{0.1 in}
As we will see, $\lambda^{*}(i, R_{1}, R_{2})$ can be interpreted as the distribution on the set of actions of the optimal i.i.d. policy that achieves $D^{*}(i, R_{1}, R_{2})$. Heuristically, a good policy would attempt to have an action process whose empirical measure on the set of actions approaches the distribution $\lambda^{*}(i, R_{1}, R_{2})$, as $\Vert \alpha \Vert \rightarrow 0$. A closed form expression for $\lambda^{*}(i, R_{1}, R_{2})$ is not available. But we now describe some structural properties of $\lambda^{*}(i, R_{1}, R_{2})$.  In particular, we show for any configuration $\Psi$, all components of $\lambda^{*}(\Psi)$ are strictly bounded away from zero.

\vspace*{0.1 in}
\begin{proposition}
\label{prop:lambda star bounded away from zero}
 Fix $K \ge 3$. Let $\lambda^{*}$ be as in (\ref{eqn: lambda star original form}). There exists a constant $c_{K} \in (0,1)$, independent of $(k, \theta_{1}, \theta_{2})$ but dependent on $K$, such that $$\lambda^{*}(k, \theta_{1}, \theta_{2})(j) > c_{K} > 0$$ for all $j \in \{1, 2, \ldots, K\}$ and for all $(k, \theta_{1}, \theta_{2})$ such that $\theta_{1} > 0 , \theta_{2} > 0$ and $\theta_{1} \ne \theta_{2}.$
\end{proposition}

\begin{IEEEproof}
 See Appendix \ref{proof: Proof of lambda bounded away from 0 and 1}.
\end{IEEEproof}
\vspace*{0.1 in}

Proposition \ref{prop:lambda star bounded away from zero}  suggests that a good policy should sample each process at least $c_{K}$ fraction of the time. Estimates of the rate of each process should then converge to the corresponding true rate. We will make use of this fact in the analysis of our proposed algorithm, which is to come shortly. 

\begin{figure}[t]
\centering
\includegraphics[scale=0.6]{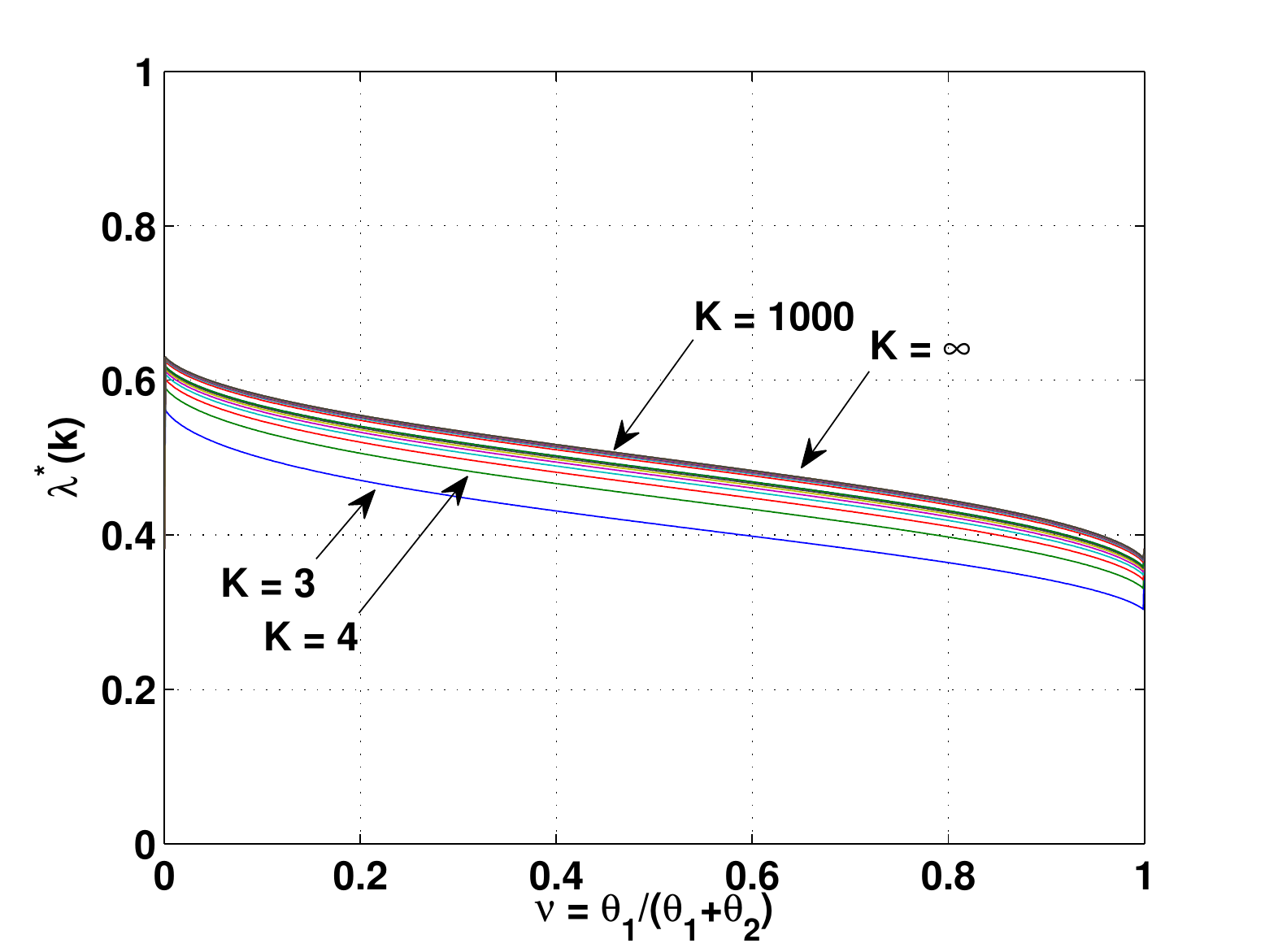}
\caption{$\lambda^{*}(k, \theta_{1}, \theta_{2})(k)$ versus $\nu = \theta_{1}/(\theta_{1}+\theta_{2})$ for various $K$.}
\label{fig:lambda vs t}
\end{figure}

An explicit expansion of the objective function in (\ref{eqn: D star simpler form}) will show that $\lambda^{*}(k, \theta_{1}, \theta_{2})(k)$ can be equivalently expressed as a function of the ratio $\nu = \theta_{1}/(\theta_{1}+\theta_{2})$. Figure \ref{fig:lambda vs t} shows the value of $\lambda^{*}(k, \theta_{1}, \theta_{2})(k)$ for different values of $\nu$ and for different $K$, $K$ varying from 3 to 1000 and $\infty$. We observe the following:
\begin{enumerate}
 \item $\lambda^{*}(k, \theta_{1}, \theta_{2})(k)$ is lower bounded by $\sim 0.3$ for all $\nu$ and for all $K$, and $\lambda^{*}(k, \theta_{1}, \theta_{2})(k)$  attains its minimum at $\nu=1$ and for $K = 3$. 
 \item $\lambda^{*}(k, \theta_{1}, \theta_{2})(k)$ is upper bounded by $\sim 0.7$ for  all $\nu$ and for all $K$, and the maximum is approached at $\nu = 0$ and as $K \rightarrow \infty$.
 \item At $\nu = 1/2$, we have $R_{1} = R_{2}$; the objective function in (\ref{eqn: D star simpler form}) is identically zero, and  any $\lambda(k)$ works. We may take $\lambda^{*}(k)$ to be the continuous extension of $\lambda^{*}(k)$ as $\nu \rightarrow 1/2$.
\end{enumerate}
From the above observations, for a fixed $K$, we have $\lambda^{*}(k, \theta_{1}, \theta_{2})(j) \gtrsim (0.3/K)$ for all $j$ and for all $(k, \theta_{1}, \theta_{2})$. In Appendix \ref{proof: Proof of lambda bounded away from 0 and 1}, where we prove Proposition \ref{prop:lambda star bounded away from zero}, we obtain a looser bound for $\lambda^{*}(k, \theta_{1}, \theta_{2})(j)$. We only show that  $\lambda^{*}(k, \theta_{1}, \theta_{2})(j) > 0.1/K$.


\section{Achievability - Modified GLRT}
\label{sec: upper bound}
 In this section we describe our proposed asymptotically optimal policy that achieves the lower bound in Proposition \ref{prop: Lower bound} as the constraint on the probability of false detection is driven to zero. Our algorithm is an adaptation of Chernoff's {\it Procedure A}. The likelihood ratio function in {\it Procedure A} is replaced by a modified generalised likelihood ratio function in our algorithm. The  strategy at each time slot is not only a function of the hypothesis with the largest GLR statistic, but also a function of the maximum likelihood estimates of the odd and non-odd rates.
 
 Before describing the algorithm, we develop some required notation.
 
 Let $N_{j}^{n}$ denote the number of times process $i$ was chosen for observation up to time $n$, i.e., $N_{j}^{n} = \sum_{t=1}^{n} 1_{\{A_{t} = j\}}$ and so $n = \sum_{j=1}^{K} N_{j}^{n}$. Let $Y_{j}^{n}$ denote the number of observed jumps in process $j$ up to time $n$; $Y_{j}^{n} = \sum_{t=1}^{n} X_{t} 1_{\{A_{t} = j\}}$. Let $Y^{n}$ denote the total number of observed jumps up to time $n$; $Y^{n} = \sum_{j=1}^{K} Y_{j}^{n}$. 
 
 Let $f(X^{n}, A^{n}\vert \Psi = (j,\theta_{1},\theta_{2}))$ be the likelihood function of the observations and actions up to time $n$, conditioned on the configuration $\Psi$, i.e.,
\begin{align}
 f(X^{n}, A^{n}\vert \Psi = (j,\theta_{1},\theta_{2})) = \frac{1}{\prod_{t=1}^{n} (X_{t}!)} ~\theta_{1}^{Y_{j}^{n}} e^{-N_{j}^{n} \theta_{1}} ~ \theta_{2}^{(Y^{n} - Y_{j}^{n})} e^{-(n-N_{j}^{n}) \theta_{2}}.  
\end{align}
Let $\beta_{11}, \beta_{12}, \beta_{21}, \beta_{22}$ be fixed constants, all greater than zero. Let
\begin{align}
 f_{\beta_{11}, \beta_{12}, \beta_{21}, \beta_{22}}(\Psi = (j, \theta_{1}, \theta_{2}) \vert H = j) &:= f_{\beta_{11}, \beta_{12}} (\theta_{1}\vert H=j) ~ f_{\beta_{21}, \beta_{22}}(\theta_{2} \vert H = j)\\
 & := \frac{\beta_{12}^{\beta_{11}} \theta_{1}^{\beta_{11}-1} e^{-\beta_{12} \theta_{1}}}{\Gamma(\beta_{11})} ~ \frac{\beta_{22}^{\beta_{21}} \theta_{2}^{\beta_{21}-1} e^{-\beta_{22} \theta_{2}}}{\Gamma(\beta_{21})} 
\end{align}
denote the product gamma densities on the parameters $\theta_{1}$ and $\theta_{2}$. The Gamma distribution is a conjugate prior for the Poisson distribution. We will use $f_{1,1,1,1}(\Psi = (j, \theta_{1}, \theta_{2}) \vert H = j)$ as an artificial prior on the parameter space $\Theta = \{(\theta_{1}, \theta_{2})\}$ in our proposed algorithm. While any positive $(\beta_{11}, \beta_{12}, \beta_{21}, \beta_{22})$ would suffice,  $(\beta_{11}, \beta_{12}, \beta_{21}, \beta_{22}) = (1,1,1,1)$ makes the calculations and the presentation simpler. $\theta_{1}$ and $\theta_{2}$ then have the exponential distribution with mean 1. 

Let $\hat{\theta}_{j}^{n} = (\hat{\theta}_{j,1}^{n}, \hat{\theta}_{j,2}^{n})$ denote the maximum likelihood estimates of the odd and non-odd rates at time $n$ conditioned on $H = j$, i.e.,
\begin{align}
 \hat{\theta}_{j,1}^{n} = \frac{Y_{j}^{n}}{N_{j}^{n}} \text{ and } \hat{\theta}_{j,2}^{n} = \frac{(Y^{n} - Y_{j}^{n})}{(n - N_{j}^{n})}.
\end{align}
Let 
\begin{align} 
 \hat{f}\left(X^{n}, A^{n} \vert H = j\right) &:= \max_{\Psi: H = j} f\left(X^{n}, A^{n}\vert \Psi\right)\\
 & = f\left(X^{n}, A^{n}\vert \Psi = (j, \hat{\theta}_{j,1}^{n}, \hat{\theta}_{j,2}^{n})\right)\\
 & = \frac{1}{\prod_{t=1}^{n} (X_{t}!)} \left(\frac{Y_{j}^{n}}{N_{j}^{n}}\right)^{Y_{j}^{n}} e^{-Y_{j}^{n}} \times \left(\frac{Y^{n} - Y_{j}^{n}}{n - N_{j}^{n}}\right)^{(Y^{n} -Y_{j}^{n})} e^{-(Y^{n} -Y_{j}^{n})}
\end{align}
denote the maximum likelihood of the observations and actions till time $n$ conditioned on $H = j$. The maximum is taken over all possible odd and non-odd rates. 
Let the averaged likelihood function at time $n$, averaged according to the artificial prior $f_{1,1,1,1}$ over all configurations $\Psi$ given $H = i$ be
\begin{align}
 f(X^{n}, A^{n} \vert H = i) &:=  \int f(X^{n}, A^{n} \vert \Psi = (i, \theta_{1}, \theta_{2})) ~ f_{1,1,1,1}((i, \theta_{1}, \theta_{2}) \vert H = i) d\theta_{1} d\theta_{2}\\
 &=   \frac{1}{\prod_{t=1}^{n} (X_{t}!)} \int \theta_{1}^{Y_{i}^{n}} e^{-N_{i}^{n} \theta_{1}} \theta_{2}^{(Y^{n} - Y_{i}^{n})} e^{-(n - N_{i}^{n}) \theta_{2}} ~ e^{-\theta_{1}} ~ e^{-\theta_{2}} d\theta_{1} d\theta_{2}\\
 \label{eqn: averaged likelihood 1} & =  \frac{1}{\prod_{t=1}^{n} (X_{t}!)} \frac{\Gamma(Y_{i}^{n}+1)}{(N_{i}^{n}+1)^{(Y_{i}^{n}+1)}} \cdot \frac{\Gamma(Y^{n} -Y_{i}^{n}+1)}{(n - N_{i}^{n}+1)^{(Y^{n} -Y_{i}^{n}+1)}},
\end{align}
where the last equality follows by recognising the presence of $\text{Gamma}(Y_{i}^{n}+1$, $N_{i}^{n}+1)$ and $ \text{Gamma}(Y^{n} -Y_{i}^{n}+1$, $n - N_{i}^{n}+1)$ densities  without scale factors  in (\ref{eqn: averaged likelihood 1}).
The modified GLR is defined as 
\begin{align}
 \label{eqn: modified GLR 1} Z_{ij}(n) &:= \log\left(\frac{f(X^{n}, A^{n} \vert H = i)}{\hat{f}\left(X^{n}, A^{n} \vert H = j\right)}\right)\\
 \nonumber & = \log \left(\frac{\Gamma(Y_{i}^{n}+1)}{(N_{i}^{n}+1)^{(Y_{i}^{n}+1)}} \cdot \frac{\Gamma(Y^{n} -Y_{i}^{n}+1)}{(n - N_{i}^{n}+1)^{(Y^{n -}Y_{i}^{n}+1)}}\right)\\
  \label{eqn: modified GLR 2} & \hspace*{1 cm}- Y_{j}^{n}\left(\log \left(\frac{Y_{j}^{n}}{N_{j}^{n}}\right) - 1\right) - (Y^{n} -Y_{j}^{n}) \left( \log \left(\frac{(Y^{n} - Y_{j}^{n})}{(n - N_{j}^{n})}\right)-1 \right).
\end{align} 
Note that the numerator is an averaged likelihood under $H = i$, averaged with respect to an artificial prior, and denominator is a maximum likelihood  under $H = j$. Let 
\begin{align}
 Z_{i}(n) := \min_{j \ne i} Z_{ij}(n)
\end{align}
 denote the modified GLR  with respect to $i$ for the nearest alternate.
 \vspace{0.1 in}
 
We now describe our proposed policy.\vspace*{0.1 in}
\begin{addmargin}[2em]{2em}
{\it Policy: {\it Modified GLRT}} ($\pi_{M}(L)$) \\
Fix $L \ge 1 $. \\
At time $n$ (end of slot $n$):
\begin{itemize}
\item Let $i^{*}(n) = \arg\max_{i} Z_{i}(n)$, the index with the largest modified GLR after $n$ time slots. Ties are resolved uniformly at random.
\item If $Z_{i^{*}(n)}(n) < \log{((K-1)L)}$ then $A_{n+1}$ is chosen according to $\lambda^{*}(i^{*}(n), \hat{\theta}_{i^{*}(n)1}^{n}, \hat{\theta}_{i^{*}(n)2}^{n})$, i.e.,
\begin{align}
\label{eqn:probability of action selection dist free}
 \Pr(A_{n+1} = j \vert X^{n}, A^{n}) = \lambda^{*}(i^{*}(n), \hat{\theta}_{i^{*}(n)1}^{n}, \hat{\theta}_{i^{*}(n)2}^{n})(j).
\end{align}
\item If $Z_{i^{*}(n)}(n)  \ge \log{((K-1)L)}$ then the test retires and declares $i^{*}(n)$ as the true hypothesis.
\end{itemize}
\end{addmargin}
\vspace*{0.1 in}
As done in \cite{ref:VaiArunSund_Journal1_Arxiv_201506}, we also consider two variants of $\pi_{M}(L)$ which are useful in the analysis.
\begin{itemize}
 \item {\it Policy} $\pi^{i}_{M}(L)$: This is the same as $\pi_{M}(L)$, but stops only at decision $i$ when $Z_{i}(n) \ge \log((K-1)L)$.
 \item {\it Policy} $\tilde{\pi}_{M}$: This is the same as $\pi_{M}(L)$, but never stops, and hence $L$ is irrelevant.
\end{itemize}
Under a fixed hypothesis $H = i$, and the triplet of policies $(\pi_{M}(L), \pi^{i}_{M}(L), \tilde{\pi}_{M}(L))$, it is easily seen that there is a common underlying probability measure with respect to which the processes $(X_n, A_n)_{n \ge 1}$ associated with the three policies are naturally coupled, with only the stopping times being different. We denote the stopping times by $ \tau(\pi_{M}(L))$ and $\tau(\pi^{i}_{M}(L))$, respectively. Under this coupling, the following are true:
\begin{align*}
 \tau(\pi^{i}_{M}(L)) & \ge \tau(\pi_{M}(L)),\\
 \{\tau (\pi_{M}(L)) > n\} &\subseteq \{\tau(\pi^{i}_{M}(L)) > n\}\\
&\subseteq \left\{Z_{i}(n) < \log((K-1)L)\right\}.
 \end{align*}

We now explore the characteristics of the proposed policy $\pi_{M}(L)$. 
\begin{proposition}
\label{prop: pi always stops}
 Fix $L > 1$. Policy $\pi_{M}(L)$ stops in finite time with probability 1, that is, $P(\tau(\pi_{M}(L)) < \infty) = 1.$
\end{proposition}
\begin{IEEEproof}
 See Appendix \ref{proof: pi always stops}.
\end{IEEEproof}

In the proof, we argue that, when the odd process has index $i$, i.e., $H = i$, the test statistic $Z_{i}(n)$ has a strictly positive drift and hence will cross the threshold $\log((K-1)L)$ in finite time almost surely. Proof is given in Appendix \ref{proof: pi always stops}.

For any $\alpha$, we show that the policy $\pi_{M}(L)$, with $L$ chosen suitably, belongs to $\Pi(\alpha)$. In other words, $\pi_{M}(L)$ satisfies the constraint on the probability of false detection.
\vspace*{0.1 in}
\begin{proposition}
\label{prop: pi in pi(alpha)}
 Fix $\alpha = (\alpha_{1}, \alpha_{2}, \ldots, \alpha_{K})$. Let $L = 1/ \min_{k}\alpha_{k}$. We then have $\pi_{M}(L) \in \Pi(\alpha).$
\end{proposition}

\begin{IEEEproof}
 From the choice of $L$, we have $1/L \le \alpha_{k}$  for all $k \in \{1, 2, \ldots, K\}$. This implies $\Pi((1/L,1/L, \ldots, 1/L)) \subseteq \Pi(\alpha)$.  Hence, it suffices to show that $\pi_{M}(L) \in \Pi((1/L,1/L, \ldots, 1/L))$. 
 
 Fix $\Psi = (i, R_{1}, R_{2})$. Let $\Delta_{j}^{n} = \{\omega: \tau(\pi_{M}(L))(\omega) = n, \delta(\omega) = j\}$ denote the sample paths for which the decision maker stops sampling after $n$ time slots and decides in favour of $H = j$. The decision region in favour of $j$ is denoted $\Delta_{j} := \cup_{n \ge 1} \Delta_{j}^{n}$. Note that 
 \begin{align}
 \label{eqn: disjoint decision regions}
  \Delta_{j}^{n} \cap \Delta_{j}^{m} = \emptyset \text{ for all } m \ne n.
 \end{align}

 We now use a standard change of measure argument to bound the conditional probability of false detection as follows, with $P$ in place of $P^{\pi_{M}}$:
 \begin{align}
  \nonumber P(\delta \ne i \vert \Psi = (i, R_{1}, R_{2})) &= \sum_{j \ne i} P(\delta = j \vert \Psi = (i, R_{1}, R_{2})) + P(\tau(\pi_{M}(L)) = \infty \vert\Psi = (i, R_{1}, R_{2}))\\
  \label{eqn: pi in pi(alpha) 1} & = \sum_{j \ne i} \sum_{n \ge 1} \int_{\omega \in \Delta_{j}^{n}} dP(\omega \vert \Psi = (i, R_{1}, R_{2}))+ 0\\
  \nonumber 			 & = \sum_{j \ne i} \sum_{n \ge 1} \int_{\omega \in \Delta_{j}^{n}} f(x^{n}, a^{n} \vert \Psi = (i, R_{1}, R_{2}))) d(x^{n}, a^{n})\\
  \label{eqn: pi in pi(alpha) 2} & \le \sum_{j \ne i} \sum_{n \ge 1} \int_{\omega \in \Delta_{j}^{n}} \left(\frac{\hat{f}(x^{n}, a^{n} \vert H = i)}{f(x^{n}, a^{n} \vert H = j)}\right) f(x^{n}, a^{n} \vert H = j) d(x^{n}, a^{n})\\
  \label{eqn: pi in pi(alpha) 3} & \le \sum_{j \ne i} \frac{1}{L(K-1)} \sum_{n \ge 1} \int_{\omega \in \Delta_{j}^{n}} f(x^{n}, a^{n} \vert H = j) d(x^{n}, a^{n})\\
  \label{eqn: pi in pi(alpha) 4} & \le \frac{1}{L}.
 \end{align}
The equality in (\ref{eqn: pi in pi(alpha) 1}) follows from (\ref{eqn: disjoint decision regions}) and from Proposition (\ref{prop: pi always stops}). The inequality in (\ref{eqn: pi in pi(alpha) 2}) follows because the maximum likelihood function satisfies $\hat{f}(x^{n}, a^{n} \vert H = i) \ge f(x^{n}, a^{n} \vert \Psi = (i, R_{1}, R_{2}))$ for all $\Psi$ such that $H = i$. The inequality in (\ref{eqn: pi in pi(alpha) 3}) follows because $\omega \in \Delta_{j}^{n}$ implies $Z_{ji} \ge \log((K-1)L)$, which in turn implies that the term within parenthesis is upper bounded by $1/((K-1)L)$, a consequence of (\ref{eqn: modified GLR 1}). Inequality in (\ref{eqn: pi in pi(alpha) 4}) follows because the inner summation in (\ref{eqn: pi in pi(alpha) 3}) is a sum of probabilities of disjoint events, and hence is upper bounded by one.
\end{IEEEproof}
\vspace*{0.1 in}
Observe that we chose the modified GLR instead of GLR precisely because we want to recognise the inner summation in (\ref{eqn: pi in pi(alpha) 3}) as a probability of an event and upper bounded by 1. If we use the  GLR,  the integrand would have been a maximum likelihood which after summation and integration may not even be finite.

We now move on to show that $\pi_{M}$ is asymptotically optimal. We first assert that the process $(Z_{i}(n))_{n \ge 1}$ has an asymptotic drift equal to $D^{*}(i, R_{1}, R_{2})$.
\vspace*{0.1 in}
\begin{proposition}
\label{prop: drift of log-likelihood ratio function}
 Consider the non-stopping policy $\tilde{\pi}_{M}$. Let $\Psi = (i, R_{1}, R_{2})$ be the true configuration. Then,
 \begin{align}
  \lim_{n \rightarrow \infty} \frac{Z_{i}(n)}{n} = D^{*}(i, R_{1}, R_{2}) ~ \text{ almost surely.} 
 \end{align}
\end{proposition}

\begin{IEEEproof}
 See Appendix \ref{proof: Zij has drift D*}.
\end{IEEEproof}
\vspace*{0.1 in}
We now state the main proposition that upper bounds the expected stopping time of our proposed policy $\pi_{M}$.
\vspace*{0.1 in}
\begin{proposition}
\label{prop: upper bound on expected stopping time}
 Consider the policy $\pi_{M}(L)$. Let $\Psi = (i, R_{1}, R_{2})$ be the true configuration. Then 
 \begin{align}
  \limsup_{L \rightarrow \infty} \frac{\tau(\pi_{M}(L))}{\log(L)} &\le \frac{1}{D^{*}(i, R_{1}, R_{2})} ~ \text{ almost surely, and further, }\\
  \limsup_{L \rightarrow \infty} \frac{E\left[\tau(\pi_{M}(L))\vert \Psi \right]}{\log(L)} &\le \frac{1}{D^{*}(i, R_{1}, R_{2})}.
  \end{align}
\end{proposition}

We now state the main theorem that combines the lower bound in Proposition \ref{prop: Lower bound}  and the upper bound in Proposition \ref{prop: upper bound on expected stopping time} to show that our proposed policy $\pi_{M}(L)$ is asymptotically optimal.
\vspace*{0.1 in}
\begin{thm}
 Consider $K$ homogeneous Poisson point processes with configuration $\Psi = (i, R_{1}, R_{2})$. Let $(\alpha^{(n)})_{n \ge 1}$ be a sequence of vectors, where $\alpha^{(n)}$ is the $n$th tolerance vector, such that  $\lim_{n \rightarrow \infty} \Vert \alpha^{(n)}\Vert = 0$ and
 \begin{align}
 \label{eqn: alpha vector max by min assumption}
  \limsup_{n \rightarrow \infty} \frac{\Vert \alpha^{(n)}\Vert}{\min_{k} \alpha_{k}^{(n)}} \le B ~ \text{ for some } B. 
 \end{align}
Then, for each $n$, the policy $\pi_{M}(L_{n})$ with $L_{n} = 1/\min_{k} \alpha_{k}^{(n)}$ belongs to $\Pi(\alpha^{(n)})$. Furthermore, 
\begin{align}
 \liminf_{n \rightarrow \infty} \inf_{\pi \in \Pi(\alpha^{(n)})} \frac{E\left[\tau(\pi))\vert \Psi\right]}{\log(L_{n})}
 & = \limsup_{n \rightarrow \infty}  \frac{E\left[\tau(\pi_{M}(L_{n})))\vert \Psi \right]}{\log(L_{n})}\\
 & = \frac{1}{D^{*}(i, R_{1}, R_{2})}.
\end{align}

\end{thm}

\begin{IEEEproof}
 The fact that $\pi_{M}(L_{n}) \in \Pi(\alpha^{(n)})$ follows from Proposition \ref{prop: pi in pi(alpha)}. We then have the following inequalities:
 \begin{align}
  \label{eqn: main theorem 1} \frac{1}{D^{*}(i, R_{1}, R_{2})} &\le \liminf_{n \rightarrow \infty} \inf_{\pi \in \Pi(\alpha^{(n)})} \frac{E\left[\tau(\pi))\vert \Psi\right]}{- \log(\Vert\alpha^{(n)}\Vert)}\\
  \label{eqn: main theorem 2} & = \liminf_{n \rightarrow \infty} \inf_{\pi \in \Pi(\alpha^{(n)})} \frac{E\left[\tau(\pi))\vert \Psi\right]}{\log(L_{n})}\\
  \label{eqn: main theorem 3} & \le \limsup_{n \rightarrow \infty} \frac{E\left[\tau(\pi_{M}(L_{n})))\vert \Psi\right]}{\log(L_{n})}\\
  \label{eqn: main theorem 4} & \le \frac{1}{D^{*}(i, R_{1}, R_{2})}.
 \end{align}
Inequality (\ref{eqn: main theorem 1}) follows from Proposition \ref{prop: Lower bound}. Equality (\ref{eqn: main theorem 2}) follows from the choice of $L_{n}$ and from assumption (\ref{eqn: alpha vector max by min assumption}). Inequality (\ref{eqn: main theorem 3}) follows because $\pi_{M}(L_{n})$ is an element in $\Pi(\alpha^{(n)})$. Inequality (\ref{eqn: main theorem 4}) follows from Proposition \ref{prop: upper bound on expected stopping time}.
\end{IEEEproof}

\section{Application to Visual Search}
\label{sec: Application to Visual Search}
In this section we apply our results to the visual search experiments of Sripati and Olson \cite{ref:201001JNS_SriOls}. A decision theoretic viewpoint of these experiments was proposed by  Vaidhiyan et al. \cite{ref:VaiArunSund_Journal1_Arxiv_201506}, and a suitable  neuronal dissimilarity index based on an ASHT model for visual search was identified. The neuronal dissimilarity index was taken as the inverse of the constant to which $E[\tau(L)/\log(L)]$ converges as $L \rightarrow \infty$, where $1/L$ is the constraint on the probability of false detection and $\tau(L)$ is the stopping time of the optimal policy. We refer the reader to Vaidhiyan et al. \cite{ref:VaiArunSund_Journal1_Arxiv_201506} for a more detailed exposition on the decision theoretic formulation. In that paper, it was assumed that $R_{1}$ and $R_{2}$ are known. If they are unknown and have to be learnt along the way, we fall into the framework of this paper, and the corresponding neuronal dissimilarity index  would be $D^{*}(i, R_{1}, R_{2})$. 

In the visual search model of Vaidhiyan et al. \cite{ref:VaiArunSund_Journal1_Arxiv_201506}, an image was assumed to elicit, in a population of neurons, a spiking pattern according to a multi-dimensional Poisson point process. Also, given the firing rates, the processes were assumed to be independent of each other. The neuronal representation for an image was then taken as the corresponding firing rate vector. Our current model accounts for a one-dimensional Poisson point process, equivalent to a single neuron scenario. However, all our results for a one-dimensional Poisson point process extends naturally  to multi-dimensional Poisson point processes. Hence, the extension of $D^{*}(i, R_{1}, R_{2})$ to $D^{*}(i, \underline{R}_{1}, \underline{R}_{2})$ for vectors of rates $\underline{R}_{1}, \underline{R}_{2}$ is straightforward - formula (\ref{eqn: D star simpler form}) continues to hold with $R_{1}$, $R_{2}$, $\tilde{R}$ replaced with vectors $\underline{R}_{1}$, $\underline{R}_{2}$, $\underline{\tilde{R}}$ respectively, and $D(R_{i},\tilde{R})$ replaced by $D(\underline{R}_{i}, \underline{\tilde{R}}) = \sum_{d} D(\underline{R}_{i}(d) \Vert \underline{\tilde{R}}(d))$, where the summation is over neuron indices.

Table \ref{table:correlation of different neuronal metrics including DF} shows the correlation values for different dissimilarity indices. See Vaidhiyan et al. \cite{ref:VaiArunSund_Journal1_Arxiv_201506} for details on the different neuronal indices and different test statistics. We see that the inverse of the proposed $D^{*}$, as with the inverse of other indices, is strongly correlated with the average decision delay. 
\begin{table}[ht]
\caption{Correlation with Different Neuronal Dissimilarity Indices} 
\centering 
\begin{tabular}{|c|c|c|} 
\hline\hline 
Information Measure & Correlation ($s$ vs $(\text{Neuronal index})^{-1}$)& $p$-value\\  
\hline 
&&\\
$\tilde{D}$ \cite{ref:VaiArunSund_Journal1_Arxiv_201506} &  0.89 & $4.3 \times 10^{-09}$\\
KL & 0.90 & $3.1 \times 10^{-09}$\\
Chernoff &  0.88 & $2.1 \times 10^{-08}$\\
$L^1$ &  0.88 & $1.1 \times 10^{-08}$\\ 
\hline
& & \\
 $D^{*}$ (this paper) & 0.89 & $8.7 \times 10^{-09}$\\
[1ex] 
\hline 
\end{tabular}
\label{table:correlation of different neuronal metrics including DF} 
\end{table}

%

An ideal neuronal dissimilarity index, say $\mathsf{diff}(i,j)$, would satisfy $E[\tau(i,j)] \mathsf{diff}(i,j) = constant$, for each image pair $(i,j)$. Vaidhiyan et al. \cite{ref:VaiArunSund_Journal1_Arxiv_201506} proposed  tests of equality of means to measure the dispersion of $E[\tau(i,j)] \mathsf{diff}(i,j)$ about a common mean. A natural statistic to test the dispersion of group means about a common mean is the ratio of arithmetic mean (AM) to geometric mean (GM) of the group means. It turns out that (AM/GM) is the statistic for a GLRT based equality of means test for Gamma distributed random variables under a fixed shape parameter assumption. The test for equality of means across groups for Gaussian random variables is the one-way ANOVA test. ANOVA is also widely used for non-Gaussian random variables also because of its robustness.

Table \ref{table:GLRT_values including DF} shows the statistics related to ANOVA and (AM/GM) tests. As with other indices, $D^{*}$ fails the equality of means tests (Indicated by the $p$-values for ANOVA in the second column. Similarly for $\log$(AM/GM) tests). When the statistics are used to rank order the indices, from the ANOVA statistic (smaller the better), we see that $D^{*}$ fares slightly worse than $\tilde{D}$, but better than the other indices. From the $\log$(AM/GM) statistics we see that $D^{*}$ performs worse than  $\tilde{D}$ and the KL indices, but better than Chernoff and $L_{1}$. 

\begin{table}[ht]
\caption{Equality of means test using various test statistics} 
\centering 
\begin{tabular}{|l|c|c|c|c|} 
\hline\hline 
 $\textsf{diff}$ & ANOVA statistic & ANOVA $p$-values &   $\log$(AM/GM)\\  
\hline 
& & &\\
 $\tilde{D}$ \cite{ref:VaiArunSund_Journal1_Arxiv_201506} & $06.30$ & $9.35 \times 10^{-19}$ & $0.0200$ \\
       KL & $06.68$ & $2.88 \times 10^{-20}$ &  $0.0211$ \\
 Chernoff & $06.74$ & $1.61 \times 10^{-20}$ &  $0.0252$ \\
  $L^{1}$ & $24.00$ & $3.42 \times 10^{-87}$ &  $0.0678$ \\
  \hline
  & & &\\
  $D^{*}$ (this paper) & $06.34$ & $6.93 \times 10^{-19}$ & $0.0233$\\
  [1ex] 
\hline 
\end{tabular}
\label{table:GLRT_values including DF} 
\end{table}


The slight degradation in performance of $D^{*}$ with respect to $\tilde{D}$ may be attributed to  the particular experimental setup of Sripati and Olson \cite{ref:201001JNS_SriOls}. The search tasks associated with a given image pair belonged to the same block of trials, and hence were contiguous. This may have possibly cued the subject about the upcoming image pair, which violates our assumption on the lack of prior knowledge of the image pairs to the decision maker. More experiments with a wide variety of image pairs and few repetitions are required for a more thorough evaluation of the performance of $D^{*}$.

\section{Conclusion}
\label{sec: Conclusion}
We studied the problem of detecting an odd Poisson point process having a rate different from the common rate of others.  We developed a lower bound on the conditional expected stopping time for any policy that satisfies the given constraint on the probability of false detection. We proposed a modified GLRT based algorithm, that we called $\pi_{M}$ and showed that it satisfies the given constraint on the probability of false detection, and that it is asymptotically optimal with respect to the conditional expected stopping time. The proposed algorithm employs  a simple threshold criterion for stopping. Interestingly, we also showed that, independent of the configuration, the sampling probability for each process is strictly above a positive constant.

We applied our results to the visual search experiments of Sripati and Olson \cite{ref:201001JNS_SriOls}. We proposed $D^{*}$ as a candidate neuronal dissimilarity index. $D^{*}$ correlated strongly with the behavioral data. The performance of $D^{*}$ was marginally inferior to the neuronal dissimilarity index proposed by Vaidhiyan et al. in \cite{ref:VaiArunSund_Journal1_Arxiv_201506}. 

This work was restricted to Poisson processes. Extension to other class of distributions, especially exponential family is under consideration. Extension to general class of distributions will be an interesting extension.

\section*{Acknowledgements}
We would like to thank Dr. S. P. Arun, from the Centre for Neuroscience, IISc Bangalore, and Prof. Carl R. Olson, from the Center for the Neural Basis of Cognition, Carnegie Mellon Unibersity, for the experimental data used in Section \ref{sec: Application to Visual Search}.


\appendices
\section{Proof of Proposition \ref{prop:lambda star bounded away from zero}}
\label{proof: Proof of lambda bounded away from 0 and 1}

 Let us rewrite (\ref{eqn: D star simpler form}) as
 \begin{align*}
  \lambda^{*}(k, \theta_{1}, \theta_{2})(k) = \arg \max_{0 \le \lambda \le 1} \left[\lambda D(\theta_{1} \Vert \tilde{\theta}) + (1 - \lambda)\frac{(K-2)}{(K-1)} D(\theta_{2} \Vert \tilde{\theta})\right],
 \end{align*}
where $\tilde{\theta}$, as in (\ref{eqn: R tilde}), is given by
\begin{align}
\label{eqn: R tilde new}
 \tilde{\theta} = \frac{\lambda \theta_{1} + (1-\lambda) \frac{(K-2)}{(K-1)} \theta_{2}}{\lambda + (1-\lambda) \frac{(K-2)}{(K-1)}}.
\end{align}
We have abused  notation and have used  $\lambda$  to denote the scalar $\lambda(k)$ of (\ref{eqn: D star simpler form}). We first show that the second derivative of the objective function in the above optimisation is negative for all $\lambda$ to establish concavity. Define the objective function as

\begin{align*}
 f(\lambda) := \lambda D(\theta_{1} \Vert \tilde{\theta}) + (1 - \lambda)\frac{(K-2)}{(K-1)} D(\theta_{2} \Vert \tilde{\theta}),
\end{align*}
where $\tilde{\theta}$, a function of $\lambda$, is as in (\ref{eqn: R tilde new}). We then have

\begin{align}
 \frac{df}{d\lambda} &=  D(\theta_{1} \Vert \tilde{\theta}) - \frac{(K-2)}{(K-1)} D(\theta_{2} \Vert \tilde{\theta}) + \left(\lambda D'(\theta_{1} \Vert \tilde{\theta}) + (1-\lambda)\frac{(K-2)}{(K-1)}D'(\theta_{2} \Vert \tilde{\theta}) \right) \frac{d \tilde{\theta}}{d \lambda}\\
 \label{eqn: derivative of f 1 new} & = D(\theta_{1} \Vert \tilde{\theta}) - \frac{(K-2)}{(K-1)} D(\theta_{2} \Vert \tilde{\theta}),
 \end{align}
where, we recall, $D'(x \Vert y)$ is the derivative of $D(x \Vert y)$ with respect to the second argument $y$, which turns out to be $1-x/y$. Equality (\ref{eqn: derivative of f 1 new}) follows from (\ref{eqn: D derivative equal to zero to obtain R}), which ensures that the term within the parenthesis is identically zero. Differentiating once again,

\begin{align}
 \frac{d^{2} f}{d \lambda^{2}}  & = \left(D'(\theta_{1} \Vert \tilde{\theta}) - \frac{(K-2)}{(K-1)} D'(\theta_{2} \Vert \tilde{\theta})\right) \frac{d \tilde{\theta}}{d \lambda}\\
  & = \left(\left(1 - \frac{\theta_{1}}{\tilde{\theta}}\right) - \frac{(K-2)}{(K-1)} \left(1 - \frac{\theta_{2}}{\tilde{\theta}}\right)\right) \frac{d \tilde{\theta}}{d \lambda}\\
  & = -\frac{\tilde{\theta}}{\lambda+(1-\lambda)\frac{(K-2)}{(K-1)}} \left(\left(1 - \frac{\theta_{1}}{\tilde{\theta}}\right) - \frac{(K-2)}{(K-1)} \left(1 - \frac{\theta_{2}}{\tilde{\theta}}\right)\right)^{2}\\
  \label{eqn: d square f by d lambda square < 0 new}& \le 0,
 \end{align}
 where we have used the fact that 
 \begin{align}
 \frac{d \tilde{\theta}}{d \lambda}  = -\frac{ \tilde{\theta}}{\lambda + (1-\lambda)\frac{(K-2)}{(K-1)}} \left(\left(1 - \frac{\theta_{1}}{\tilde{\theta}}\right) - \frac{(K-2)}{(K-1)} \left(1 - \frac{\theta_{2}}{\tilde{\theta}}\right)\right).
\end{align}
Since $f(\lambda)$ is concave in $\lambda$, and since $f(0) = f(1) = 0$, and $f'(0) > 0$ and $f'(1) < 0$,  the maximiser $\lambda^{*}$ satisfies
\begin{align}
\label{eqn: optimum lambda derivative relation new}
  D(\theta_{1} \Vert \tilde{\theta}) - \frac{(K-2)}{(K-1)} D(\theta_{2} \Vert \tilde{\theta}) = 0.
\end{align}
We do not know of a closed form expression for $\lambda^{*}$ from (\ref{eqn: optimum lambda derivative relation new}). 

\begin{figure}[t]
\centering
\includegraphics[scale=0.6]{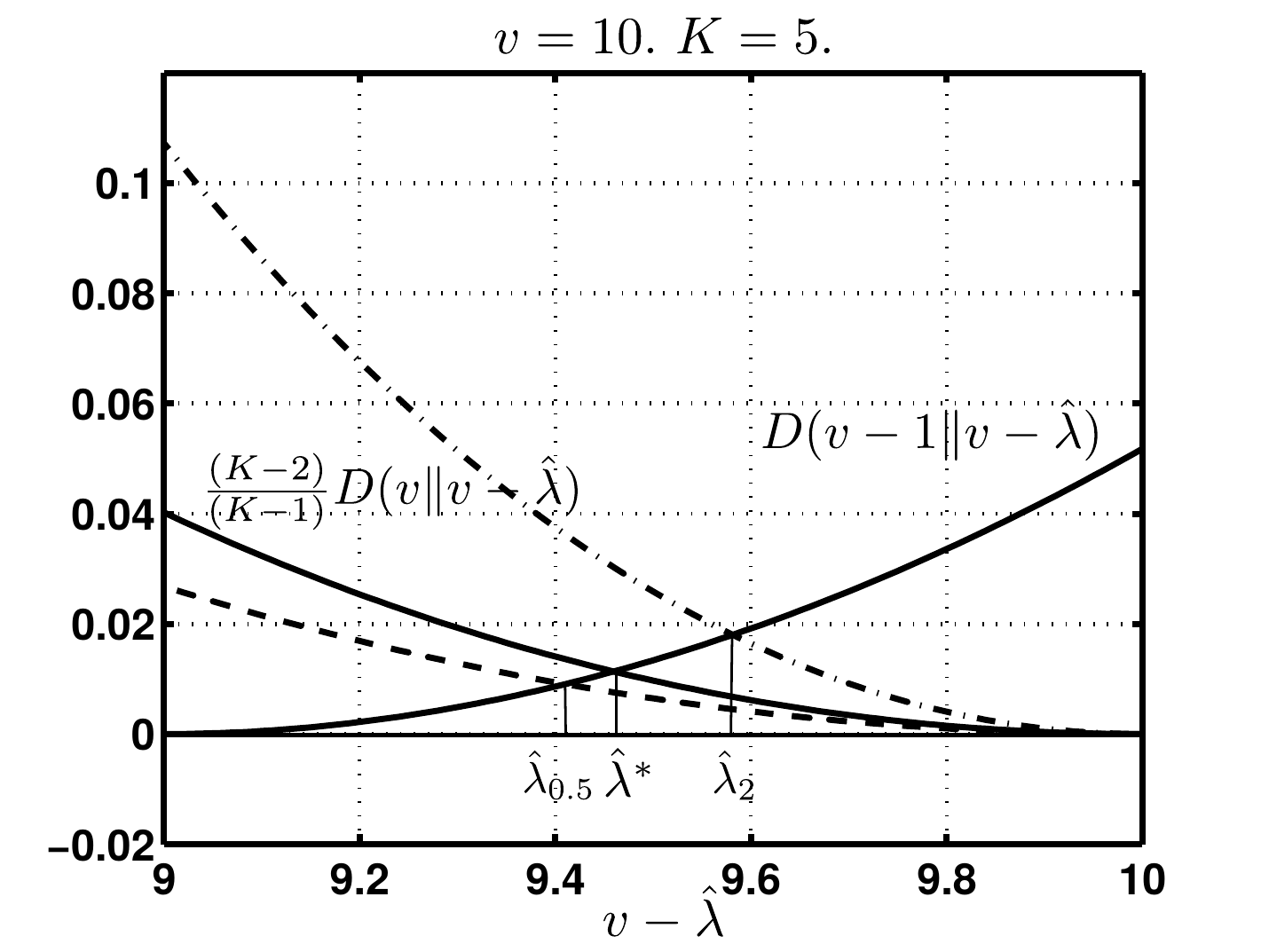}
\caption{Geometric interpretation of $\hat{\lambda}^{*}$.}
\label{fig: geometric interpretation lambda hat star UB and LB}
\end{figure}

Let $\hat{\lambda}$ denote a parametrisation of $\lambda$ of the form
\begin{align}
 \label{eqn:lambda hat}
 \hat{\lambda} := \frac{\lambda}{\lambda+(1-\lambda) \frac{(K-2)}{(K-1)}},
\end{align}
so that $\tilde{\theta} = \hat{\lambda} \theta_{1} + (1-\hat{\lambda}) \theta_{2}$. We recognise that $\hat{\lambda}$ is increasing in $\lambda$. Let $\hat{\lambda}^{*}$ denote the re-parametrisation for $\lambda^{*}$ according to (\ref{eqn:lambda hat}). Hence, to show that $\lambda^{*}$ is bounded away from 0 and 1, it suffices to show that $\hat{\lambda}^{*}$ is bounded away from 0 and 1. Let us first consider the case when $\theta_{1} < \theta_{2}$. The case when $\theta_{1} > \theta_{2}$ has similar arguments. Let us consider a new parametrisation of (\ref{eqn: optimum lambda derivative relation new}). Let $v$ denote 
\begin{align}
 v = \frac{\theta_{2}}{\theta_{2}-\theta_{1}},
\end{align}
so that 
\begin{align}
 v-1 = \frac{\theta_{1}}{\theta_{2}-\theta_{1}},
\end{align}
and 
\begin{align}
 \frac{\tilde{\theta}}{\theta_{2}-\theta_{1}} &= \frac{\hat{\lambda} \theta_{1}+(1-\hat{\lambda}) \theta_{2}}{\theta_{2}-\theta_{1}}\\
 & = v-\hat{\lambda}.
\end{align}
The left-hand side of (\ref{eqn: optimum lambda derivative relation new}) can now be written in terms of $v$ and $\hat{\lambda}$ as
\begin{align}
\label{eqn: D in terms of v and lambda}
 D(v-1 \Vert v-\hat{\lambda}) - \frac{(K-2)}{(K-1)} D(v \Vert v - \hat{\lambda}).
\end{align}
Let $\hat{\lambda}_{r}(v)$ denote the solution to 
\begin{align}
 D(v-1 \Vert v-\hat{\lambda}) - r D(v \Vert v - \hat{\lambda}) = 0.
\end{align}  
Figure \ref{fig: geometric interpretation lambda hat star UB and LB} gives a geometric interpretation of $\hat{\lambda}^{*}$. Note that $\hat{\lambda}^*(v) = \hat{\lambda}_r(v)$ for $r = (K-2)/(K-1)$. For each $v \geq 1$, we also have that $\hat{\lambda}_r(v)$ decreases with $r$. Furthermore, $0.5 \leq (K-2)/(K-1) \leq 2$. We then have $\hat{\lambda}_{2}(v) < \hat{\lambda}^{*}(v) < \hat{\lambda}_{0.5}(v)$. Hence, to show that $\hat{\lambda}^{*}(v)$  is bounded away from 0 and 1 for all $v$, it suffices to show that $\sup_{v \geq 1} \hat{\lambda}_{0.5}(v) < 1$, and that $\inf_{v \ge 1} \hat{\lambda}_{2}(v) > 0$. 

We now obtain a Taylor's series based alternate expression for $D(v-a \Vert v-b)$ when $v \ge 1$ and $0 \le a,b \le 1$. The alternate expression replaces the log terms in (\ref{eqn: D in terms of v and lambda}) with infinite sums and enables easier bounding of (\ref{eqn: D in terms of v and lambda}).

\begin{lemma}
 \label{lemma: new expression for KL divergence}
 Let $v \ge 1$. Let $0 \le a,b \le 1$. Let $D(x \Vert y) = x\log(x/y)-y+x$ denote the relative entropy between two Poisson random variables with means $x$ and $y$. Then,
 \begin{align}
  D(v-a \Vert v-b) &= (v-a) \log\left(\frac{v-a}{v-b}\right)-(v-a)+(v-b)\\
  &=\sum_{l \ge 1} \frac{1}{v^{l}l(l+1)} \left(a^{l+1} - b^{l}(a+(a-b)l)\right).
 \end{align}
\end{lemma}

\begin{figure}[t]
\centering
\includegraphics[scale=0.6]{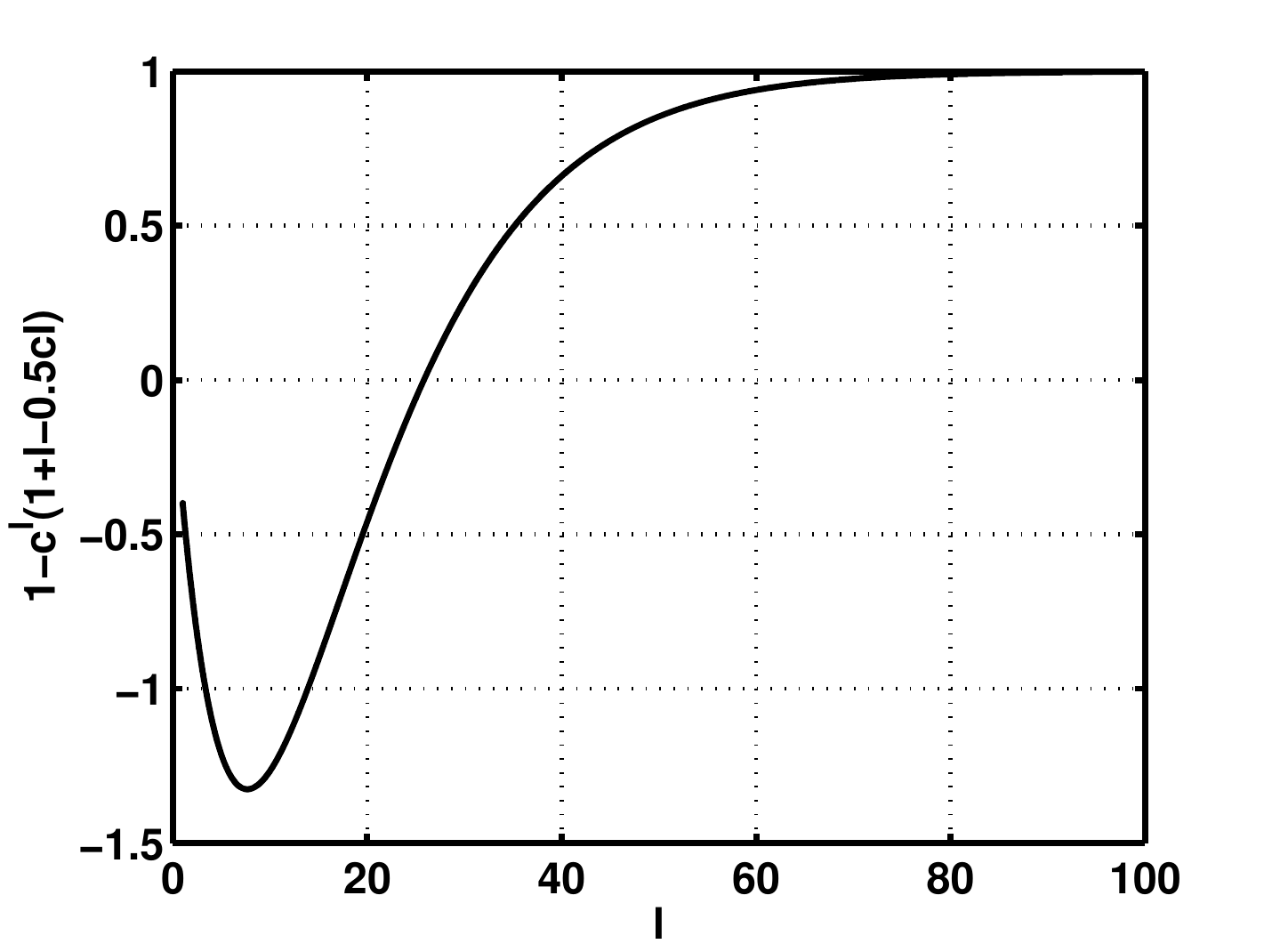}
\caption{Variation of $(1-c^{l}(1+l-0.5cl))$ with $l$.}
\label{fig: negative then positive in l}
\end{figure}

\begin{IEEEproof}
 \label{proof: lemma - new expression for KL divergence}
 Case 1: Let $v > 1$. Let $0 \le a,b \le 1$.\\
 Using the Taylor's series expansion for $-\log(1-x) = \sum_{l \ge 1} \frac{x^{1}}{l}$, when $|x| < 1$, we get
 \begin{align}
   D(v-a \Vert v-b) &= (v-a) \log\left(\frac{1-a/v}{1-b/v}\right)-(v-a)+(v-b)\\
   & = -(v-a) \sum_{l \ge 1} \frac{a^{l}}{v^{l}l}+(v-a)\sum_{l \ge 1}\frac{b^{l}}{v^{l}l} +(a-b)\\
   & = (-a+b)+ \sum_{l \ge 2} \frac{1}{v^{l-1}l} \left(b^{l}-a^{l}\right) + \sum_{l \ge 1}  \frac{1}{v^{l}l} \left(a^{l+1}-ab^{l}\right) + (a-b)\\
   & = \sum_{l \ge 1} \frac{1}{v^{l}(l+1)} \left(b^{l+1}-a^{l+1}\right) + \sum_{l \ge 1}  \frac{1}{v^{l}l} \left(a^{l+1}-ab^{l}\right)\\
   & = \sum_{l \ge 1} \frac{1}{v^{l}l(l+1)} \left(a^{l+1}-b^{l}(a+(a-b)l)\right).
 \end{align}
 Case 2: Let $v = 1$. Let $0 < a,b < 1$. The same arguments as above holds.\\
 Case 3: Let $v = 1$. Let $a=1$, $b < 1$. Then, 
  \begin{align}
   \sum_{l \ge 1} \frac{1}{l(l+1)} \left(1-b^{l}(1+(1-b)l)\right) &=  \sum_{l \ge 1}\left[ \frac{1}{l}- \frac{1}{(l+1)}- \frac{b^{l}}{l} + \frac{b^{l+1}}{l+1} \right] \\
   &= (1-b)\\
   & = D(0 \Vert 1-b).
 \end{align}
 Case 4: Let $v=1$. Let $a < 1$, $b = 1$. Then, both  $D(v-a \Vert v-b)$ and the infinite sum are infinity.\\
 Case 5: Let $v=1$. Let $a=1, b=1$. Then both $D(v-a \Vert v-b)$ and the infinite sum are zero.
\end{IEEEproof}
\vspace*{0.1 in}
We now show that $\hat{\lambda}_{0.5}(v) < 0.9$  for all $v \ge 1$. For this, it suffices to show that for $c = 0.9$, $D(v-1 \Vert v-c) - 0.5 D(v \Vert v-c) < 0$ for all $v \ge 1$.
\begin{align}
 D(v-1 \Vert v-c) -0.5 D(v \Vert v-c) &= \sum_{l \ge 1} \frac{1}{v^{l}l(l+1)} \left(1 - c^{l}(1+(1-c)l) - 0.5 c^{l+1}l \right)\\
 & = \sum_{l \ge 1} \frac{1}{v^{l}l(l+1)} \left(1 - c^{l}(l+1) + c^{l+1}l - 0.5 c^{l+1}l \right)\\
 & = \sum_{l \ge 1} \frac{1}{v^{l}l(l+1)} \left(1 - c^{l}(l+1 - 0.5 cl \right).
\end{align}
Let us first consider the case when $v=1$ and $c = 0.9$. We then have
\begin{align}
 \nonumber D(v-1 \Vert v-c) -0.5 D(v \Vert v-c) &= D(0\Vert0.1)-0.5 D(1 \Vert 0.1)\\
 \nonumber & = 0.1 - 0.5(\log(10)-0.9)\\
\label{eqn: lambda hat 0.5 < 0.9 for v = 1} & < 0.
\end{align}
Thus, $\hat{\lambda}_{0.5}(1) < 0.9$. For $v > 1$ and $c = 0.9$, we observe that $(1-c^{l}(1+l-0.5cl))$ is initially negative and then becomes positive in $l$ (See Figure \ref{fig: negative then positive in l}). Thus, there exists $M >1$ such that 
\begin{align}
 (1-c^{l}(1+l-0.5cl))
 \begin{cases}
  \le 0 \quad \forall~l < M\\
  \ge 0 \quad \forall~l \ge M.
 \end{cases}
\end{align}
Then, for $c = 0.9$, we  have
\begin{align}
 \nonumber D(v-1 \Vert v-c) -0.5 D(v \Vert v-c) &= \sum_{l \ge 1} \frac{1}{v^{l}l(l+1)} \left(1 - c^{l}((l+1) - 0.5 cl \right)\\
   \label{eqn: lambda hat for any v eqn 2} & \le \sum_{1 \le l < M} \frac{1}{v^{M}l(l+1)} \left(1 - c^{l}((l+1) - 0.5 cl \right) + \sum_{l \ge M} \frac{1}{v^{M}l(l+1)} \left(1 - c^{l}((l+1) - 0.5 cl \right)\\
 \nonumber & = \frac{1}{v^{M}} \sum_{l \ge 1} \frac{1}{l(l+1)} \left(1 - c^{l}((l+1) - 0.5 cl \right)\\
 \nonumber & = \frac{1}{v^{M}} \left(D(0 \Vert 1-c) -0.5 D(1 \Vert 1-c)\right)\\
 \nonumber & = \frac{1}{v^{M}} \left(D(0 \Vert 0.1) -0.5 D(1 \Vert 0.1)\right)\\
 \label{eqn: lambda hat for any v eqn 6} & < 0.
\end{align}
Inequality (\ref{eqn: lambda hat for any v eqn 2}) is obtained  by upperbounding 1) the initial negative terms, till $l < M$, by replacing $v^{l}$ by a larger $v^{M}$, and 2) the later non-negative terms, for $l \ge M$, by replacing $v^{l}$ by a smaller $v^{M}$.  Inequality (\ref{eqn: lambda hat for any v eqn 6}) follows from (\ref{eqn: lambda hat 0.5 < 0.9 for v = 1}).
Thus, we have shown that $\hat{\lambda}_{0.5}(v) < 0.9$ for all $v \ge 1$.

We now show the second part of the proof, i.e., $\hat{\lambda}_{2}(v) > 0.1>0$. For this, it suffices to show that for $c= 0.1$, $D(v-1 \Vert v-c) - 2 D(v \Vert v-c) > 0$ for all $v \ge 1$. For $c = 0.1$, we have 
\begin{align}
 \label{eqn:c0.1 eqn 1} D(v-1 \Vert v-c) - 2 D(v \Vert v-c) &= \sum_{l \ge 1} \frac{1}{v^{l}l(l+1)} \left(1-c^{l}(1+(1-c)l) - 2 c^{l+1}l\right)\\
 \label{eqn:c0.1 eqn 2}& =  \sum_{l \ge 1} \frac{1}{v^{l}l(l+1)} \left(1-c^{l}(1+l+cl)\right)\\
 \label{eqn:c0.1 eqn 3} & =  \sum_{l \ge 1} \frac{1}{v^{l}l(l+1)} \left(1-(0.1)^{l}(1+l+(0.1)l)\right)\\
 \label{eqn:c0.1 eqn 4} & > 0,
\end{align}
where (\ref{eqn:c0.1 eqn 4}) follows as each term inside the summation in (\ref{eqn:c0.1 eqn 3}) is positive.
Thus, when $\theta_{1} < \theta_{2}$ and for all $v \ge 1$, we have shown that  
\begin{align}
 0.1 \le \hat{\lambda}_{2}(v) \le \hat{\lambda^{*}}(v) \le  \hat{\lambda}_{0.5}(v) < 0.9.
\end{align}

We now consider the case when $\theta_{1} > \theta_{2}$. Let 
\begin{align}
 v' = \frac{\theta_{1}}{\theta_{1} - \theta_{2}},
\end{align}
so that 
\begin{align}
 v' -1 = \frac{\theta_{2}}{\theta_{1} - \theta_{2}},
\end{align}
and 
\begin{align}
\frac{\tilde{\theta}}{\theta_{1}-\theta_{2}} &= \frac{\hat{\lambda} \theta_{1}+(1-\hat{\lambda}) \theta_{2}}{\theta_{1}-\theta_{2}}\\
 & = v'-1+\hat{\lambda}.
\end{align}

Equation (\ref{eqn: optimum lambda derivative relation new}) can now be written in terms of $v'$ and $\hat{\lambda}$ as
\begin{align}
\label{eqn: D in terms of v_prime and lambda hat}
 D(v'\Vert v'-1+\hat{\lambda}) - \frac{(K-2)}{(K-1)} D(v'-1 \Vert v' -1 + \hat{\lambda}) = 0.
\end{align}
Let $\hat{\lambda^{*}}(v')$ be the solution to (\ref{eqn: D in terms of v_prime and lambda hat}). Recognise that (\ref{eqn: D in terms of v_prime and lambda hat}) has the same form as in the previous case for $\theta_{1} < \theta_{2}$, with only the multiplicative constant being different. From arguments similar to the ones used in the previous case of $\theta_{1} < \theta_{2}$, we can show that $$0.1 < (1-\hat{\lambda^{*}}(v'))<0.9,$$ or equivalently, $0.1 < \hat{\lambda}^{*}(v') < 0.9$.

Thus, we have shown that $\hat{\lambda}^{*}$ is bounded away from 0.1 and 0.9 for all $\theta_{1}$ and $\theta_{2}$.
\hfill \IEEEQEDclosed

\section{}
\label{appendix: Properties of policy pi MGLRT}
We stated the main properties of the proposed policy $\pi_{M}$ in Section \ref{sec: upper bound}. We prove them in this Appendix.  
\subsection{Proof of Proposition \ref{prop: pi always stops} and associated ingredients}
\label{proof: pi always stops}
Before we prove Proposition \ref{prop: pi always stops} we develop some convergence results for $\pi_{M}(L)$. We show that under the non-stopping policy $\tilde{\pi}_{M}$, the empirical rate associated with a process converges to the true rate of that process. The results are akin to convergence results for independent random variables, but applied to the dependent random variables in our setting with the dependency being induced by the policy.
\vspace*{0.1 in}
\begin{proposition}
\label{prop: empirical rate convergence}
 Fix $K \ge 3$. Let $\Psi = (i, R_{1}, R_{2})$ be the true configuration. Consider the non stopping policy $\tilde{\pi}_{M}$. As $n \rightarrow \infty$ the following convergences hold almost surely,
 \begin{align}
 \label{eqn: convergence of empirical rates}
  \frac{Y_{j}^{n}}{N_{j}^{n}} & \rightarrow 
  \begin{cases}
   R_{1} & \text{ if } j = i,\\
   R_{2} & \text{ if } j \ne i,
  \end{cases}
  \end{align},
  \begin{align}
   \frac{Y^{n} - Y_{i}^{n}}{n-N_{i}^{n}} \rightarrow R_{2},
  \end{align}
  and
  \begin{align}
  \label {eqn: Yn - Yjn statement}
  R'_{min} \le \liminf_{n \rightarrow \infty} \frac{Y^{n} - Y_{j}^{n}}{n-N_{j}^{n}} \le \limsup_{n \rightarrow \infty} \frac{Y^{n} - Y_{j}^{n}}{n-N_{j}^{n}}\le  R'_{max} \text{ for all } j \ne i,
 \end{align}
where 
\begin{align}
 \label{eqn: R' min}
 R'_{min} = (1-c_{K}) \min\{R_{1}, R_{2}\} + c_{K} \max\{R_{1}, R_{2}\}
\end{align}
and
\begin{align}
 \label{eqn: R' max}
 R'_{max} = c_{K} \min\{R_{1}, R_{2}\} + (1-c_{K}) \max\{R_{1}, R_{2}\}, 
\end{align}
and $c_{K}$ is as in Proposition \ref{prop:lambda star bounded away from zero}.
\end{proposition}

\begin{IEEEproof}
 Let  $\mathcal{F}_{l-1}$ denote the $\sigma$-field generated by $(X^{l-1}, A^{l-1})$. Consider the martingale difference sequence $$S_{i}^{n} = Y_{i}^{n} - N_{i}^{n} R_{1} = \sum_{l=1}^{n} (X_{l}-R_{1}) 1_{\{A_{l} = i\}}.$$ Given the Poisson assumption on $X_{l}$, we have $E\left[(X_{l}-R_{1})^{2} 1_{\{A_{l} = i\}} \vert \mathcal{F}_{l-1}\right] < \infty$ for all $l$. Then, by the convergence result for martingales, see De la Pena \cite[Theorem 1.2A]{ref:victor1999general}, for any $\epsilon > 0$, there exists $c_{\epsilon} > 0$ such that 
 \begin{align}
 \label{eqn: De la Pena martingale concentration result}
  P(S_{i}^{n}>n \epsilon) \le e^{-c_{\epsilon} n},
 \end{align}
 which in turn, by the Borel-Cantelli Lemma \cite[sec 4.2]{book:chung2001course}, implies
 \begin{align}
 \label{eqn: martingale convergence 1}
  \frac{S_{i}^{n}}{n} \rightarrow 0 \text{ almost surely.}
 \end{align}
 Similarly arguing, we conclude that convergence result holds for other $S_{j}^{n} / n$, for $ j = 1,2, \ldots, K$.
 Further, from Proposition \ref{prop:lambda star bounded away from zero}, we have 
 \begin{align}
 \label{eqn: gamma i greater than 0}
  \liminf_{n \rightarrow \infty} \frac{N_{i}^{n}}{n} > c_{K} > 0 \text{ almost surely.}
 \end{align}
Combining (\ref{eqn: martingale convergence 1}) and (\ref{eqn: gamma i greater than 0}), we have, 
\begin{align}
 \frac{S_{i}^{n}}{N_{i}^{n}} \rightarrow 0 \text{ almost surely,}
 \end{align}
 or equivalently,
 \begin{align}
 \frac{Y_{i}^{n}}{N_{i}^{n}} \rightarrow R_{1} \text{ almost surely.}
\end{align}
Similar result hold for other $j$, with $R_{1}$ replaced by $R_{2}$, and we have established  (\ref{eqn: convergence of empirical rates}). Furthermore, these results imply that
\begin{align}
\label{eqn: Yn-Yjn}
 \frac{(Y^{n} - Y_{j}^{n}) - \sum_{k \ne j} N_{k}^{n} R_{k}}{(n-N_{j}^{n})} & \rightarrow 0 \text{ almost surely.}
\end{align}
Consequently, we get 
\begin{align}
\label{eqn: Yn-Yin convergence}
 \frac{(Y^{n} - Y_{i}^{n})}{(n-N_{i}^{n})} & \rightarrow R_{2} \text{ almost surely.}
\end{align}
Fix $j \ne i$,  we then have 
\begin{align}
 \frac{\sum_{k \ne j} N_{k}^{n} R_{k}}{(n-N_{j}^{n})}  = \frac{N_{i}^{n}}{n-N_{j}^{n}} R_{1} + \frac{\sum_{k \ne j,i} N_{k}^{n} }{n-N_{j}^{n}}R_{2}.
\end{align}
We do not yet have a convergence result for $N_{k}^{n}/n$ for any $k$. Proposition (\ref{prop:lambda star bounded away from zero}) only says that at every slot and for each process,  the probability of choosing that process is greater than $c_{K}$. Thus, we are not in a position to say, as $n \rightarrow \infty$, whether  $$\frac{Y^{n} - Y_{j}^{n}}{n-N_{j}^{n}} \rightarrow constant.$$ However, from Proposition \ref{prop:lambda star bounded away from zero}, we get the following bound 
\begin{align}
\nonumber (1-c_{K}) \min \{R_{1}, R_{2}\} + c_{K} \max \{R_{1}, R_{2}\} & \le \liminf_{n \rightarrow \infty}\frac{\sum_{k \ne j} N_{k}^{n} R_{k}}{(n-N_{j}^{n})} \le \limsup_{n \rightarrow \infty}\frac{\sum_{k \ne j} N_{k}^{n} R_{k}}{(n-N_{j}^{n})}\\
 \label{eqn: Rmin Rmax interval}& \le c_{K} \min\{R_{1}, R_{2}\} + (1-c_{K}) \max\{R_{1}, R_{2}\} \text{ almost surely}.
\end{align}
Thus, (\ref{eqn: Yn-Yjn}) combined with (\ref{eqn: Rmin Rmax interval}) yields (\ref{eqn: Yn - Yjn statement}).
\end{IEEEproof}
\vspace*{0.1 in}

We now state a lemma that asserts that, under the non-stopping policy $\tilde{\pi}_{M}$, $Z_{i}(n)$, the test statistic associated with the index of the odd process, drifts off to infinity.
\begin{lemma}
\label{lemma: Zij by n greater than zero}
 Fix $K \ge 3$. Let $\Psi = (i, R_{1}, R_{2})$ be the true configuration. Consider the non-stopping policy $\tilde{\pi}_{M}$. Then for all $j \ne i$, we have
 \begin{align}
  \liminf_{n \rightarrow \infty} \frac{Z_{ij}(n)}{n} > 0 \text{ almost surely.}
 \end{align}
\end{lemma}

 \begin{IEEEproof}
 Without loss of generality assume $R_{1} < R_{2}$. Observe that we have $R_{1} < R'_{min} < R'_{max} < R_{2}$. Recall that $$D(x \Vert y) := x\log(x/y) -x +y,$$ the relative entropy between two Poisson distributions with means $x$ and $y$. We can write (\ref{eqn: modified GLR 2}) as
 
  \begin{align}
   \nonumber Z_{ij}(n) & = \log \left(\frac{\Gamma(Y_{i}^{n}+1)}{(N_{i}^{n}+1)^{(Y_{i}^{n}+1)}}   \frac{\Gamma(Y^{n} -Y_{i}^{n}+1)}{(n - N_{i}^{n}+1)^{(Y^{n -}Y_{i}^{n}+1)}}\right)\\ 
   \nonumber & \hspace*{1 cm} - Y_{j}^{n}\left(\log \left(\frac{Y_{j}^{n}}{N_{j}^{n}}\right) - 1\right) - (Y^{n} -Y_{j}^{n}) \left( \log \left(\frac{Y^{n} - Y_{j}^{n}}{n - N_{j}^{n}}\right)-1 \right)\\
   \nonumber & \ge  Y_{i}^{n} \log \left(\frac{Y_{i}^{n}}{N_{i}^{n}+1}\right) - Y_{i}^{n} + \log \left(\frac{\sqrt{2\pi Y_{i}^{n}}}{N_{i}^{n}+1}\right)\\
   \nonumber & \hspace{1 cm} + (Y^{n}-Y_{i}^{n})\log \left(\frac{Y^{n} - Y_{i}^{n}}{n - N_{i}^{n}+1}\right) - (Y^{n}-Y_{i}^{n})+ \log \left(\frac{\sqrt{2\pi (Y^{n}-Y_{i}^{n})}}{n - N_{i}^{n}+1}\right)\\
   \label{eqn: Zij lower bound 1} & \hspace{1 cm} - \left[ Y_{j}^{n}\log \left(\frac{Y_{j}^{n}}{N_{j}^{n}}\right) -  Y_{j}^{n} + (Y^{n} -Y_{j}^{n})  \log \left(\frac{Y^{n} - Y_{j}^{n}}{n - N_{j}^{n}}\right)- (Y^{n} -Y_{j}^{n}) \right]\\
  \nonumber & = (N_{i}^{n}+1) D\left(\frac{Y_{i}^{n}}{N_{i}^{n}+1} \Vert \frac{Y^{n}-Y_{j}^{n}}{n-N_{j}^{n}}\right)+ (n - N_{i}^{n}-N_{j}^{n}) D\left(\frac{Y^{n} - Y_{i}^{n} - Y_{j}^{n}}{n - N_{i}^{n}-N_{j}^{n}} \Vert \frac{Y^{n}-Y_{j}^{n}}{n-N_{j}^{n}}\right)\\
  \nonumber &\hspace{1 cm} -N_{j}^{n} D\left(\frac{Y_{j}^{n}}{N_{j}^{n}} \Vert \frac{Y^{n}-Y_{i}^{n}}{n-N_{i}^{n}+1}\right) - (n - N_{i}^{n}-N_{j}^{n}) D\left(\frac{Y^{n} - Y_{i}^{n} - Y_{j}^{n}}{n - N_{i}^{n}-N_{j}^{n}} \Vert \frac{Y^{n}-Y_{i}^{n}}{n-N_{i}^{n}+1}\right)\\
  \label{eqn: Zij lower bound 4}& \hspace{1 cm} - \frac{Y^{n}-Y_{i}^{n}}{n-N_{i}^{n}+1} - \frac{Y^{n} - Y_{j}^{n}}{n - N_{j}^{n}}+ \log \left(\frac{\sqrt{2 \pi Y_{i}^{n}}}{N_{i}^{n}+1}\right) + \log \left(\frac{\sqrt{2 \pi (Y^{n} - Y_{i}^{n}})}{n-N_{i}^{n}+1}\right),
  \end{align}
 where the inequality (\ref{eqn: Zij lower bound 1}) follows from the lower bound for the gamma function $\Gamma(x+1) = x! \ge x^{x} e^{-x} \sqrt{2 \pi x}$ \cite[p.54]{book:Feller_AnIntroductionToProbTheory_1968}, and the equality (\ref{eqn: Zij lower bound 4}) follows from the use of the formula for $D(x \Vert y)$ and some rearrangement of terms.
 
 We now study the convergence of each of the terms in (\ref{eqn: Zij lower bound 4}). All convergence statements are in the almost sure sense. Consider the first term in (\ref{eqn: Zij lower bound 4}). From Proposition \ref{prop: empirical rate convergence} and Proposition \ref{prop:lambda star bounded away from zero}, as $n \rightarrow \infty$, we have
 \begin{align}
  \frac{Y_{i}^{n}}{N_{i}^{n}+1} \rightarrow R_{1} \text{, } \liminf_{n \rightarrow \infty} \frac{Y^{n}-Y_{j}^{n}}{n - N_{j}^{n}} \ge R'_{min} \text {, and } \liminf_{n \rightarrow \infty} \frac{N_{i}^{n}+1}{n} \ge c_{K}.
 \end{align} 
Consequently, and using the fact that $D(x \Vert y)$ is monotone increasing in $y$, for $y > x$, we have 
\begin{align}
 \liminf_{n \rightarrow \infty} \frac{N_{i}^{n}+1}{n} D\left(\frac{Y_{i}^{n}}{N_{i}^{n}+1} \Vert \frac{Y^{n}-Y_{j}^{n}}{n-N_{j}^{n}}\right) \ge c_{K} D\left(R_{1} \Vert R'_{min}\right) > 0.
\end{align}
Similarly, for the the second term in (\ref{eqn: Zij lower bound 4}), we have
 \begin{align}
  \frac{Y^{n}-Y_{i}^{n}-Y_{j}^{n}}{n-N_{i}^{n}-N_{j}^{n}} \rightarrow R_{2} \text{,  } \limsup_{n \rightarrow \infty} \frac{Y^{n}-Y_{j}^{n}}{n - N_{j}^{n}} \le R'_{max} \text{, and } \liminf_{n \rightarrow \infty} \frac{n-N_{i}^{n} - N_{j}^{n}}{n} \ge (K-2)c_{K} \ge c_{K}.
 \end{align}
 Consequently, and using the fact that $D(x \Vert y)$ is monotone decreasing in $y$, for $y < x$, we have
\begin{align}
 \liminf_{n \rightarrow \infty} \frac{n-N_{i}^{n} - N_{j}^{n}}{n} D\left(\frac{Y^{n} - Y_{i}^{n} - Y_{j}^{n}}{n - N_{i}^{n}-N_{j}^{n}} \Vert \frac{Y^{n}-Y_{j}^{n}}{n-N_{j}^{n}}\right) \ge c_{K} D(R_{2} \Vert R'_{max}) > 0.
\end{align}
Consider the third term in (\ref{eqn: Zij lower bound 4}). From Proposition \ref{prop: empirical rate convergence}, as $n \rightarrow \infty$, we have
 \begin{align}
 \frac{Y_{j}^{n}}{N_{j}^{n}} \rightarrow R_{2} \text{ and } \frac{Y^{n}-Y_{i}^{n}}{n - N_{i}^{n}+1}\rightarrow R_{2}.
 \end{align}
Consequently,
\begin{align}
D\left(\frac{Y^{n} - Y_{i}^{n} - Y_{j}^{n}}{n - N_{i}^{n}-N_{j}^{n}} \Vert \frac{Y^{n}-Y_{j}^{n}}{n-N_{j}^{n}}\right) \rightarrow D(R_{2} \Vert R_{2}) = 0.
\end{align}
Similarly, for the fourth term in (\ref{eqn: Zij lower bound 4}) we get
 \begin{align}
 \frac{Y^{n}- Y_{i}^{n}-Y_{j}^{n}}{n - N_{i}^{n}-N_{j}^{n}} \rightarrow R_{2} \text{ and } \frac{Y^{n}-Y_{i}^{n}}{n - N_{i}^{n}+1}\rightarrow R_{2}.
 \end{align}
Consequently,
\begin{align}
D\left(\frac{Y^{n} - Y_{i}^{n} - Y_{j}^{n}}{n - N_{i}^{n}-N_{j}^{n}} \Vert \frac{Y^{n}-Y_{i}^{n}}{n-N_{i}^{n}+1}\right) \rightarrow D(R_{2} \Vert R_{2}) = 0.
\end{align}
 Consider the fifth and sixth terms in (\ref{eqn: Zij lower bound 4}). From Proposition \ref{prop: empirical rate convergence}, we have
  \begin{align}
 \frac{Y^{n}-Y_{i}^{n}}{n-N_{i}^{n}+1} \rightarrow R_{2} \text{ and } \limsup_{n \rightarrow \infty} \frac{Y^{n} - Y_{j}^{n}}{n - N_{j}^{n}} \le R'_{max}.
 \end{align}
Consequently, when divided by $n$ and as $n \rightarrow \infty$, both the terms go to zero, i.e.,
\begin{align}
 \frac{1}{n}\frac{Y^{n}-Y_{i}^{n}}{n-N_{i}^{n}+1} \rightarrow 0 \text{ and } \limsup_{n \rightarrow \infty} \frac{1}{n} \frac{Y^{n} - Y_{j}^{n}}{n - N_{j}^{n}} = 0.
\end{align}
Consider the seventh and eight terms in (\ref{eqn: Zij lower bound 4}). Both the terms go to negative infinity, but only logarithmically in $n$, and hence when divided by $n$ and as $n \rightarrow \infty$, we get
\begin{align}
 \frac{1}{n} \log \left(\frac{\sqrt{2 \pi Y_{i}^{n}}}{N_{i}^{n}+1}\right) \rightarrow 0 \text{ and } \frac{1}{n} \log \left(\frac{\sqrt{2 \pi (Y^{n} - Y_{i}^{n}})}{n-N_{i}^{n}+1}\right) \rightarrow 0.
\end{align}
Thus, we have
 \begin{align}
    \label{eqn: Zij lower bound 2} \liminf_{n \rightarrow \infty} \frac{Z_{ij}(n)}{n} & \ge \liminf_{n \rightarrow \infty} ~ \left[\frac{(N_{i}^{n}+1)}{n} D\left(\frac{Y_{i}^{n}}{N_{i}^{n}+1} \Vert \frac{Y^{n}-Y_{j}^{n}}{n-N_{j}^{n}}\right)+ \frac{(n - N_{i}^{n}-N_{j}^{n})}{n} D\left(\frac{Y^{n} - Y_{i}^{n} - Y_{j}^{n}}{n - N_{i}^{n}-N_{j}^{n}} \Vert \frac{Y^{n}-Y_{j}^{n}}{n-N_{j}^{n}}\right)\right]\\
  \label{eqn: Zij lower bound 3} & \ge c_{K} D(R_{1}\Vert R'_{min}) + c_{K} D(R_{2} \Vert R'_{max})\\
  \nonumber & > 0.
 \end{align}
 This completes the proof of Lemma \ref{lemma: Zij by n greater than zero}.
 \end{IEEEproof}
\vspace*{0.1 in}

\begin{IEEEproof}[Proof of Proposition \ref{prop: pi always stops}]
We now have the ingredients to prove Proposition \ref{prop: pi always stops}. The following inequalities hold almost surely,
\begin{align}
 \nonumber \tau(\pi_{M}(L)) & \le \tau(\pi^{i}_{M}(L))\\
 \nonumber & = \inf \{n \ge 1 \vert Z_{i}(n) > \log((K-1)L)\}\\
 \nonumber & \le \inf \{ n \ge 1 \vert Z_{ij}(n') > \log((K-1)L) \text{ for all } n' \ge n \text{ and for all } j \ne i\}\\
 & < \infty,
\end{align}
where the last inequality follows from Lemma \ref{lemma: Zij by n greater than zero}.
\end{IEEEproof}
\vspace*{0.1 in}

While in Proposition \ref{prop: empirical rate convergence} we established that under the non-stopping policy $\tilde{\pi}_{M}$ $Y_{k}^{n} / N_{k}^{n} \rightarrow R_{k}$ almost surely,  the question of convergence of $N_{k}^{n} /n$ to some real constant under the $\tilde{\pi}_{M}$ policy remained to be established. We now show that under the $\tilde{\pi}_{M}$ policy it does converge to a real constant. Furthermore, we show that $(Y^{n} - Y_{j}^{n}) / (n-N_{j}^{n})$ also converges to a constant.
\vspace*{0.1 in}
\begin{proposition}
\label{prop: convergence of R tilde}
 Fix $K \ge 3$. Let $\Psi = (i, R_{1}, R_{2})$ be the true configuration. Consider the non-stopping policy $\tilde{\pi}_{M}$. Then as $n \rightarrow \infty$, the following convergences hold almost surely,
 \begin{enumerate}[(i)]
 \item \begin{align}
  \label{eqn: i star converges to i} i^{*}(n) & \rightarrow i,
  \end{align}
   \item
 \begin{align}
 \label{eqn: theta star 1 converges to R1} \hat{\theta}_{i^{*}(n),1}^{n} &\rightarrow R_{1},
  \end{align}
   \item
 \begin{align}
 \label{eqn: theta star 2 converges to R2} \hat{\theta}_{i^{*}(n),2}^{n} &\rightarrow R_{2},
  \end{align}
   \item
 \begin{align}
 \label{eqn: lambda n converges to lambda star} \lambda^{*}(i^{*}(n),\hat{\theta}_{i^{*}(n),1}^{n},\hat{\theta}_{i^{*}(n),2}^{n}) &\rightarrow \lambda^{*}(i, R_{1}, R_{2}),
  \end{align}
   \item
 \begin{align}
  \label{eqn: Njn converges to lambda star j} \frac{N_{j}^{n}}{n} &\rightarrow \lambda^{*}(i, R_{1}, R_{2})(j) \text{ for all } j = 1,2, \ldots, K,
  \end{align}
   \item
 \begin{align}
 \label{eqn: Yn-Yjn converges to Rmix} \frac{Y^{n}-Y_{j}^{n}}{n - N_{j}^{n}} & \rightarrow \tilde{R}(\lambda^{*}(i, R_{1}, R_{2})(i)) \text{ for all } j\ne i,
  \end{align}  
 \end{enumerate}
 where $\tilde{R}$ is as in (\ref{eqn: R tilde}).
\end{proposition}

\begin{IEEEproof}
 From Lemma \ref{lemma: Zij by n greater than zero} we have
 \begin{align}
  \liminf_{n \rightarrow \infty} Z_{i}(n) = \liminf_{n \rightarrow \infty} \min_{j \ne i} Z_{ij}(n) > 0 \text{ almost surely.}
 \end{align}
Fix $j \ne i$. Then, the following inequalities hold almost surely,
\begin{align*}
 \limsup_{n \rightarrow \infty} Z_{j}(n) & = \limsup_{n \rightarrow \infty} \min_{k \ne j} Z_{jk}(n)\\
 & \le \limsup_{n \rightarrow \infty} Z_{ji}(n)\\
 & \le \limsup_{n \rightarrow \infty} -Z_{ij}(n)\\
 & \le - \liminf_{n \rightarrow \infty} \min_{k \ne i} Z_{ik}(n)\\
 & = - \liminf_{n \rightarrow \infty} Z_{i}(n)\\
 & < 0.
\end{align*}
It further implies, $i^{*}(n) = \max_{k} Z_{k}(n) = i$ almost surely. This proves (i).\\
All convergence statements are in the almost sure sense.  From (i) and Proposition \ref{prop: empirical rate convergence} we get
\begin{align*}
 \hat{\theta}_{i^{*}(n),1}^{n} = \frac{Y_{i^{*}(n)}^{n}}{N_{i^{*}(n)}^{n}} \rightarrow \frac{Y_{i}^{n}}{N_{i}^{n}} \rightarrow R_{1},
\end{align*}
and similarly we get,
\begin{align*}
 \hat{\theta}_{i^{*}(n),2} = \frac{Y^{n} - Y_{i^{*}(n)}^{n}}{n - N_{i^{*}(n)}^{n}} \rightarrow \frac{Y^{n} -Y_{i}^{n}}{n - N_{i}^{n}} \rightarrow R_{2}.
\end{align*}
This proves (ii) and (iii).\\
From (i), (ii) and (iii) we have
\begin{align*}
 \lambda^{*}(i^{*}(n),\hat{\theta}_{i^{*}(n),1}^{n},\hat{\theta}_{i^{*}(n),2}^{n}) & \rightarrow \lambda^{*}(i,\hat{\theta}_{i,1}^{n},\hat{\theta}_{i,2}^{n})\\
 & \rightarrow \lambda^{*}(i,R_{1}, R_{2}),
\end{align*}
where we have used that fact that $\lambda^{*}(i,x,y)$ is jointly continuous in $(x,y)$, a fact that follows from Berge's Maximum Theorem \cite{ref:LawRay_GeneralisedMaxThm_1993}.\\
Consider the martingale sequence $N_{j}^{n} - \sum_{l = 1}^{n}  \lambda^{*}(i^{*}(n),\hat{\theta}_{i^{*}(n),1}^{n},\hat{\theta}_{i^{*}(n),2}^{n})(j)$. From (iv) and  martingale convergence arguments, as used in (\ref{eqn: martingale convergence 1}), we get 
\begin{align*}
 \frac{N_{j}^{n}}{n} & \rightarrow  \frac{1}{n} \sum_{l =1}^{n}\lambda^{*}(i^{*}(n),\hat{\theta}_{i^{*}(n),1}^{n},\hat{\theta}_{i^{*}(n),2}^{n})(j)\\
 & \rightarrow \lambda^{*}(i, R_{1}, R_{2})(j).
\end{align*}
For ease of notation, let $\lambda^{*}(i)$ denote $\lambda^{*}(i, R_{1}, R_{2})(i)$. We can rewrite $(Y^{n} - Y_{j}^{n}) / (n - N_{j}^{n})$ as
\begin{align*}
 \frac{Y^{n} - Y_{j}^{n}}{n - N_{j}^{n}} &= \frac{Y_{i}^{n} + Y^{n} - Y_{i}^{n} - Y_{j}^{n}}{n - N_{j}^{n}}\\
 & = \left[{\frac{N_{i}^{n}}{n} \frac{Y_{i}^{n}}{N_{i}^{n}} + \frac{n - N_{i}^{n} - N_{j}^{n}}{n} \frac{Y^{n} - Y_{i}^{n} - Y_{j}^{n}}{n-N_{i}^{n}-N_{j}^{n}} }\right] {\frac{n}{n-N_{j}^{n}}}.
\end{align*}
Then, from (v) we have the following convergence in almost sure sense,
\begin{align*}
 \frac{Y^{n} - Y_{j}^{n}}{n - N_{j}^{n}} &\rightarrow \frac{\lambda^{*}(i) R_{1} + (1-\lambda^{*}(i)) \frac{(K-2)}{(K-1)}R_{2}}{\lambda^{*}(i) + (1-\lambda^{*}(i)) \frac{(K-2)}{(K-1)}}\\
 & = \tilde{R}(\lambda^{*}(i)). 
\end{align*}
This completes the proof of the Proposition.
\end{IEEEproof}

\subsection{Proof of Proposition \ref{prop: drift of log-likelihood ratio function}}
\label{proof: Zij has drift D*}
 We already established (\ref{eqn: Zij lower bound 4}). Using Proposition \ref{prop: convergence of R tilde}, we now recognise that all the fractions converge to their respective quantities. Hence, 
 \begin{align}
  \label{eqn: Zij convergence 1} \liminf_{n \rightarrow \infty} \frac{Z_{ij}(n)}{n} & \ge \liminf_{n \rightarrow \infty} ~ \left[\frac{(N_{i}^{n}+1)}{n} D\left(\frac{Y_{i}^{n}}{N_{i}^{n}+1} \Vert \frac{Y^{n}-Y_{j}^{n}}{n-N_{j}^{n}}\right)+ \frac{(n - N_{i}^{n}-N_{j}^{n})}{n} D\left(\frac{Y^{n} - Y_{i}^{n} - Y_{j}^{n}}{n - N_{i}^{n}-N_{j}^{n}} \Vert \frac{Y^{n}-Y_{j}^{n}}{n-N_{j}^{n}}\right)\right]\\
  \label{eqn: Zij convergence 2} & = (\lambda^{*}(i, R_{1}, R_{2})(i)) D(R_{1} \Vert \tilde{R}) + (1 - (\lambda^{*}(i, R_{1}, R_{2})(i)))\frac{(K-2)}{(K-1)} D(R_{2} \Vert \tilde{R})\\
  \label{eqn: Zij convergence 3} & = D^{*}(i, R_{1}, R_{2}) \quad \text{ almost surely}.
 \end{align}
 Similarly, by using $\Gamma(x+1) = x! \le x^{x} e^{-x+1} \sqrt{2 \pi x}$, and following the steps leading to (\ref{eqn: Zij lower bound 4}) with limsup instead of liminf, it can be shown that $\limsup_{n \rightarrow \infty} \frac{Z_{ij}(n)}{n} \le D^{*}(i, R_{1}, R_{2})$  almost surely. It follows that 
 \begin{align*}
  \lim_{n \rightarrow \infty} \frac{Z_{i}(n)}{n} =  D^{*}(i, R_{1}, R_{2}) \text{ almost surely},
 \end{align*}
which establishes Proposition \ref{prop: drift of log-likelihood ratio function}.
\hfill \QEDclosed \\

From Proposition \ref{prop: Lower bound} we know that the expected stopping time, $E\left[\tau(\pi_{M}(L))\right]$,  grows to infinity as $L \rightarrow \infty$, but we now show that $\tau(\pi_{M}(L))$ grows to infinity in almost sure sense also.
\vspace*{0.1 in}
\begin{lemma}
 \label{lemma: stopping time increases with L}
 Fix $K \ge 3$. Let $\Psi = (i, R_{1}, R_{2})$ be the true configuration. Consider the policy $\pi_{M}(L)$. Then,
 \begin{align}
  \liminf_{L \rightarrow \infty} \tau(\pi_{M}(L)) \rightarrow \infty \text{ almost surely.}
 \end{align}
\end{lemma}

\begin{IEEEproof}
 It is evident that the sequence of random variables  $\tau(\pi_{M}(L))$, indexed by $L$, is non-decreasing in $L$. Hence, it suffices to show that, as $L \rightarrow \infty$,
 \begin{align}
  P\left(\tau(\pi_{M}(L)) < n \right) \rightarrow 0 \text{ for all } n. 
 \end{align}
 To see this, observe that
 \begin{align}
  \nonumber \limsup_{L \rightarrow \infty }P\left(\tau(\pi_{M}(L)) < n  \right) & = \limsup_{L \rightarrow \infty} P\left(\max_{1 \le l \le n} Z_{j}(l) > \log((K-1)L) \text{ for some }j  \right)\\
  \label{eqn: tau goes to infinity 1} & \le  \limsup_{L \rightarrow \infty } \sum_{j=1}^{K} \sum_{l=1}^{n}P\left(Z_{j}(l) > \log((K-1)L)\right)\\
  \label{eqn: tau goes to infinity 2} & \le  \limsup_{L \rightarrow \infty } \frac{1}{\log((K-1)L)} \sum_{j=1}^{K} \sum_{l=1}^{n} E\left[l + 2 (Y^{l})^{2}\right]\\
  \label{eqn: tau goes to infinity 3} & \le  \limsup_{L \rightarrow \infty } \frac{1}{\log((K-1)L)} \sum_{j=1}^{K} \sum_{l=1}^{n} \left[l + 2 l^{2} (\max\{R_{1},R_{2}\} + (\max \{R_{1},R_{2}\})^{2})\right]\\
  \nonumber & = 0.
 \end{align}
 Inequality (\ref{eqn: tau goes to infinity 1}) follows from union bound. In inequality (\ref{eqn: tau goes to infinity 3}) we have used the convexity of $x^{2}$ to bound $E[(\sum_{k = 1}^{l} X_{k})^{2}] < l^{2} E[(X_{k})^{2}]$, and also that for Poisson random variables $E[X^{2}] = E[X] + E[X]^{2}$. Inequality (\ref{eqn: tau goes to infinity 2}) is obtained by bounding $Z_{j}(l)$ as follows:
 \begin{align}
  \nonumber Z_{j}(l) & = \log \left(\frac{f(X^{l},A^{l}\vert H= j)}{\max_{k \ne j} \hat{f}(X^{l},A^{l}\vert H= i)}\right)\\
  \label{eqn: Zj(l) upper bounding 1} &  \le \log \left(\frac{\hat{f}(X^{l},A^{l}\vert H= j)}{\hat{f}(X^{l},A^{l}\vert H= k)}\right) \text{ for some } k \ne j\\
  \label{eqn: Zj(l) upper bounding 2} &  = Y_{j}^{l} \log\left(\frac{Y_{j}^{l}}{N_{j}^{l}}\right)-Y_{j}^{l} + (Y^{l} - Y_{j}^{l}) \log\left(\frac{Y^{l} -Y_{j}^{l}}{l -N_{j}^{l}}\right)-(Y^{l} -Y_{j}^{l})\\
  \label{eqn: Zj(l) upper bounding 3} & \hspace{2cm} -\left[Y_{k}^{l} \log\left(\frac{Y_{k}^{l}}{N_{k}^{l}}\right)-Y_{k}^{l} + (Y^{l} - Y_{k}^{l}) \log\left(\frac{Y^{l} -Y_{k}^{l}}{l -N_{k}^{l}}\right)-(Y^{l} -Y_{k}^{l})\right]\\
  \label{eqn: Zj(l) upper bounding 4} & \le (Y_{j}^{l})^{2}+ (Y^{l} -Y_{j}^{l})^{2}+ l - \left[N_{k}^{l} D\left(\frac{Y_{k}^{l}}{N_{k}^{l}} \Vert 1\right)+ (l - N_{k}^{l}) D\left(\frac{Y^{l} -Y_{k}^{l}}{l -N_{k}^{l}} \Vert 1\right) \right]\\
  \label{eqn: Zj(l) upper bounding 5} & \le (Y_{j}^{l})^{2}+ (Y^{l} -Y_{j}^{l})^{2}+ l\\
  \label{eqn: Zj(l) upper bounding 6} & \le 2 (Y^{l})^{2}+l.
 \end{align}

Inequality (\ref{eqn: Zj(l) upper bounding 1}) follows by upper bounding the numerator in by the maximum likelihood function and lower bounding the denominator by choosing the maximum likelihood function with respect to an arbitrary $k \ne j$ instead of the maximiser. Inequality (\ref{eqn: Zj(l) upper bounding 4}) follows by recognising that the terms inside square brackets in (\ref{eqn: Zj(l) upper bounding 3}) can be written as a sum of relative entropy terms minus an $l$. Also, we upper bound $x\log(x/N) -x$ by $x^{2}$. Inequality (\ref{eqn: Zj(l) upper bounding 5}) follows by ignoring the negative terms. Inequality (\ref{eqn: Zj(l) upper bounding 6}) follows by upper bounding $Y_{j}^{l}$ and $Y^{l} - Y_{j}^{l}$ by $Y^{l}$. 
\end{IEEEproof}
\vspace*{0.1 in}

In Proposition \ref{prop: drift of log-likelihood ratio function} we showed that, as $n \rightarrow \infty$ and under the non-stopping policy $\tilde{\pi}_{M}$, $Z_{i}(n)/n$ converges to $D^{*}(i, R_{1}, R_{2})$. We now show that, as $L \rightarrow \infty$,  $Z_{i}(\tau(\pi_{M}(L))) / \tau(\pi_{M}(L)) \rightarrow D^{*}(i, R_{1}, R_{2})$.
\vspace*{0.1 in}
\begin{lemma}
 \label{lemma: Zi(tau) by tau also converges to D*}
 Fix $K \ge 3$. Let $\Psi = (i, R_{1}, R_{2})$ be the true configuration. Consider the policy $\pi_{M}(L)$. We then have
 \begin{align}
 \label{eqn: Zi(tau) by tau also converges to D*}
  \lim_{L \rightarrow \infty} \frac{Z_{i}(\tau(\pi_{M}(L)))}{\tau(\pi_{M}(L))} = D^{*}(i,R_{1}, R_{2}) \text{ almost surely.}
 \end{align} 
\end{lemma}

\begin{IEEEproof}
 It follows from Proposition \ref{prop: drift of log-likelihood ratio function} and Lemma \ref{lemma: stopping time increases with L}.
\end{IEEEproof}

\subsection{Proof of Proposition \ref{prop: upper bound on expected stopping time}}
\label{proof: proof of upper bound on expected stopping time}
We now have all the ingredients to prove the main achievability result of Proposition \ref{prop: upper bound on expected stopping time}. By the definition of $\tau(\pi_{M}(L))$, we have that $Z_{i}(\tau(\pi_{M}(L))-1) < \log((K-1)L)$ at the previous slot. Using this we get
\begin{align}
 \label{eqn: achievability 1} \limsup_{L \rightarrow \infty} \frac{Z_{i}(\tau(\pi_{M}(L))-1)}{\log(L)} & \le  \limsup_{L \rightarrow \infty} \frac{\log((K-1)L)}{\log L}.
\end{align}
Substituting (\ref{eqn: Zi(tau) by tau also converges to D*}) in (\ref{eqn: achievability 1}), we get
\begin{align}
 \limsup_{L \rightarrow \infty} \frac{\tau(\pi_{M}(L))}{\log(L)} & =  \limsup_{L \rightarrow \infty} \frac{\tau(\pi_{M}(L)-1)}{\log(L)}\\
 & \le \frac{1}{D^{*}(i, R_{1}, R_{2})}.
\end{align}
%
%
A sufficient condition to establish convergence of the expected stopping time is to show that $$\limsup_{L \rightarrow \infty} E\left[\exp\left(\frac{\tau(\pi_{M}(L))}{\log(L)}\right)\right] < \infty.$$
Without loss of generality assume $R_{1} < R_{2}$, such that $R_{1} < R'_{min} < R'_{max} < R_{2}$, where $R'_{min}$ and $R'_{max}$ are as defined in (\ref{eqn: R' min}) and (\ref{eqn: R' max}), respectively. Let $\epsilon > 0$ be an arbitrary constant. Let $c_{K}$ be as in Proposition \ref{prop:lambda star bounded away from zero}. We then have
\begin{align}
 \limsup_{L \rightarrow \infty} E\left[e^{\frac{\tau(\pi_{M}(L))}{\log(L)}}\right] & = \limsup_{L \rightarrow \infty} \int_{x \ge 0} P\left(\frac{\tau(\pi_{M}(L))}{\log(L)} > \log(x) \right) dx\\
\label{eqn: uniform integrability 1} &  \le \limsup_{L \rightarrow \infty} \int_{x \ge 0} P\left(\tau^{i}(\pi_{M}(L)) > \lfloor \log(x) \log(L)\rfloor \right) dx.
 \end{align}
 Let us now define 
 \begin{align}
  \label{eqn: definition of u(L)} u(L):= \exp\left(\frac{3(1+\epsilon) \log((K-1)L)}{c_{K} D(R_{1} \Vert R'_{min}) \log(L)} + \frac{1}{\log(L)} \right).
 \end{align}
For $x < u(L)$ let us upper bound the probability by 1. We then get the right-hand side of (\ref{eqn: uniform integrability 1}) to be
\begin{align}
 \limsup_{L \rightarrow \infty} \int_{x \ge 0} P &\left(\tau^{i}(\pi_{M}(L)) > \lfloor \log(x) \log(L)\rfloor \right) dx \\
 \label{eqn: uniform integrability 2} & \le \limsup_{L \rightarrow \infty} \left[u(L) + \int_{x \ge u(L)} P\left(\tau^{i}(\pi_{M}(L)) > \lfloor \log(x) \log(L)\rfloor \right) dx\right].
 \end{align}
 Recognising that $P\left(\tau^{i}(\pi_{M}(L)) > \lfloor \log(x) \log(L)\rfloor \right)$ is constant in the interval $$x \in \left[ \exp\left(\frac{n}{\log(L)}\right),  \exp\left(\frac{n+1}{\log(L)}\right) \right)$$ and recognising that the interval length is upper bounded by  $\exp\left(\frac{n+1}{\log(L)}\right)$, we can further upper bound (\ref{eqn: uniform integrability 2}) by
 \begin{align}
 \limsup_{L \rightarrow \infty} \int_{x \ge 0} P &\left(\tau^{i}(\pi_{M}(L)) > \lfloor \log(x) \log(L)\rfloor \right) dx \\
 \label{eqn: uniform integrability 3} & \le \exp\left({\frac{3(1+\epsilon)}{c_{K} D(R_{1} \Vert R'_{min})}}\right)+ \limsup_{L \rightarrow \infty} \sum_{n \ge \lfloor {\log(u(L)) \log(L)}\rfloor} \exp\left({\frac{n+1}{\log(L)}}\right) P\left(\tau^{i}(\pi_{M}(L)) > n \right) dx\\
  \label{eqn: uniform integrability 4} & \le \exp\left({\frac{3(1+\epsilon)}{c_{K} D(R_{1} \Vert R'_{min})}}\right)+ \limsup_{L \rightarrow \infty} \sum_{n \ge \lfloor {\log(u(L)) \log(L)}\rfloor} \exp\left({\frac{n+1}{\log(L)}}\right) P\left(Z_{i}(n) < \log((K-1)L) \right) dx.
 \end{align} 
To show that the right-hand side of (\ref{eqn: uniform integrability 4}) is finite, it suffices to show that for all $$n \ge \lfloor {\log(u(L)) \log(L)}\rfloor \ge \frac{3(1+\epsilon)\log((K-1)L)}{c_{K} D(R_{1} \Vert R'_{min})}$$ and for sufficiently large $L$, there exist constants $\gamma > 0$ and  $0 < B < \infty$ such that
\begin{align}
 P\left(Z_{i}(n) < \log((K-1)L)\right) < B e^{-\gamma n}.
\end{align}
 
 We now show that such an exponential bound does exist. 
 \begin{lemma}
  \label{lemma: exponential bound for Zi(n)}
  Fix $K \ge 3$. Fix $L > 1$. Let $\Psi = (i, R_{1}, R_{2})$ be the true configuration. Let $u(L)$ be as in (\ref{eqn: definition of u(L)}). Then, there exist constants $\gamma > 0$ and $0 < B < \infty$, independent of $L$, such that for all $n \ge \lfloor u(L) \log(L)\rfloor$, we have
  \begin{align}
   P\left(Z_{i}(n) < \log((K-1)L)\right) \le B e^{-\gamma n}.
  \end{align}
 \end{lemma}

\begin{IEEEproof}
The following upper bounds for  $P\left(Z_{i}(n) < \log((K-1)L)\right)$ is self evident 
\begin{align*}
 P & \left(Z_{i}(n) < \log((K-1)L))\right)\\
 & = P\left(\min_{j \ne i} Z_{ij}(n) < \log((K-1)L)\right)\\
 & \le \sum_{j \ne i} P\left(Z_{ij}(n) < \log((K-1)L)\right).
 \end{align*}
 It now suffices to show that for every $j \ne i$ the probability term in the above expression is exponentially bounded. We upper bound $Z_{ij}(n)$ in the same way as we earlier did in (\ref{eqn: Zij lower bound 4}).  
 \begin{align}
 \nonumber P &\left(Z_{ij}(n) \le \log((K-1)L)\right)\\
 \nonumber & \le  P \left( (N_{i}^{n}+1) D\left(\frac{Y_{i}^{n}}{N_{i}^{n}+1} \Vert \frac{Y^{n}-Y_{j}^{n}}{n-N_{j}^{n}}\right)+ (n - N_{i}^{n}-N_{j}^{n}) D\left(\frac{Y^{n} - Y_{i}^{n} - Y_{j}^{n}}{n - N_{i}^{n}-N_{j}^{n}} \Vert \frac{Y^{n}-Y_{j}^{n}}{n-N_{j}^{n}}\right)\right.\\
  \nonumber &\hspace{1 cm} -N_{j}^{n} D\left(\frac{Y_{j}^{n}}{N_{j}^{n}} \Vert \frac{Y^{n}-Y_{i}^{n}}{n-N_{i}^{n}+1}\right) - (n - N_{i}^{n}-N_{j}^{n}) D\left(\frac{Y^{n} - Y_{i}^{n} - Y_{j}^{n}}{n - N_{i}^{n}-N_{j}^{n}} \Vert \frac{Y^{n}-Y_{i}^{n}}{n-N_{i}^{n}+1}\right)\\
  \label{eqn: Zij decays exponentially 1}&  \hspace{1 cm} \left. - \frac{Y^{n}-Y_{i}^{n}}{n-N_{i}^{n}+1} - \frac{Y^{n} - Y_{j}^{n}}{n - N_{j}^{n}}+ \log \left(\frac{\sqrt{2 \pi Y_{i}^{n}}}{N_{i}^{n}+1}\right) + \log \left(\frac{\sqrt{2 \pi (Y^{n} - Y_{i}^{n}})}{n-N_{i}^{n}+1}\right) < \log((K-1)L) \right)
 \end{align}
 Using union bound, we upper bound (\ref{eqn: Zij decays exponentially 1}) by a sum of probability terms as given next. 
 \begin{align}
 \nonumber P &\left(Z_{ij}(n) \le \log((K-1)L)\right)\\
 \nonumber & \le P \left( (N_{i}^{n}+1) \left(D\left(\frac{Y_{i}^{n}}{N_{i}^{n}+1} \Vert \frac{Y^{n}-Y_{j}^{n}}{n-N_{j}^{n}}\right)- D(R_{1} \Vert R'_{min})\right) < -\epsilon' n \right)\\
  \nonumber &\hspace{1 cm} + P \left((n - N_{i}^{n}-N_{j}^{n}) D\left(\frac{Y^{n} - Y_{i}^{n} - Y_{j}^{n}}{n - N_{i}^{n}-N_{j}^{n}} \Vert \frac{Y^{n}-Y_{j}^{n}}{n-N_{j}^{n}}\right) < - \epsilon' n\right)\\
  \nonumber &\hspace{1 cm} + P \left( -N_{j}^{n} D\left(\frac{Y_{j}^{n}}{N_{j}^{n}} \Vert \frac{Y^{n}-Y_{i}^{n}}{n-N_{i}^{n}+1}\right) < -\epsilon' n\right)\\
  \nonumber &\hspace{1 cm} + P \left( - (n - N_{i}^{n}-N_{j}^{n}) D\left(\frac{Y^{n} - Y_{i}^{n} - Y_{j}^{n}}{n - N_{i}^{n}-N_{j}^{n}} \Vert \frac{Y^{n}-Y_{i}^{n}}{n-N_{i}^{n}+1}\right) < -\epsilon' n\right)\\
  \nonumber &\hspace{1 cm} + P \left( - \frac{Y^{n}-Y_{i}^{n}}{n-N_{i}^{n}+1} <  -\epsilon' n\right) +  P \left( - \frac{Y^{n} - Y_{j}^{n}}{n - N_{j}^{n}} < -\epsilon' n\right) \\
  \nonumber &\hspace{1 cm} + P \left(\log \left(\frac{\sqrt{2 \pi Y_{i}^{n}}}{N_{i}^{n}+1}\right) < -\epsilon' n\right)  + P \left( \log \left(\frac{\sqrt{2 \pi (Y^{n} - Y_{i}^{n}})}{n-N_{i}^{n}+1}\right) < -\epsilon' n\right)\\
  \label{eqn: exponential bound 1} &\hspace{1 cm} + P \left( (N_{i}^{n}+1) D(R_{1} \Vert R'_{min}) - 8 \epsilon' n < \log((K-1)L) \right).
\end{align}
Let us choose $0 < \epsilon'' < c_{K}/3$, so that
\begin{align}
\label{eqn: choice of epsilon''}
 \frac{c_{K}}{1-c_{K}(1-\epsilon'')} > c_{K}(1+\epsilon '').
\end{align}
We then we choose $\epsilon' > 0$ such that $$3(1+\epsilon)(c_{K}(1-\epsilon'') D(R_{1} \Vert R'_{min}) - 8 \epsilon') > c_{K} D(R_{1} \Vert R'_{min}),$$ so that
 \begin{align}
 \label{eqn: last term equal zero}
  P \left( (N_{i}^{n}+1) D(R_{1} \Vert R'_{min}) - 8 \epsilon' n < \log((K-1)L), (N_{i}^{n}+1) > c_{K}(1-\epsilon '') n \right) = 0
 \end{align}
for all $n$ under consideration, i.e., for all $$n \ge \lfloor {\log(u(L)) \log(L)}\rfloor \ge \frac{3(1+\epsilon)\log((K-1)L)}{c_{K} D(R_{1} \Vert R'_{min})}.$$ The last term in (\ref{eqn: exponential bound 1}) can then be upper bounded by
\begin{align}
  \nonumber P & \left( (N_{i}^{n}+1) D(R_{1} \Vert R'_{min}) - 8 \epsilon' n < \log((K-1)L) \right) \\
   \nonumber & \le P \left( (N_{i}^{n}+1) D(R_{1} \Vert R'_{min}) - 8 \epsilon' n < \log((K-1)L), (N_{i}^{n}+1) > c_{K}(1-\epsilon '') n \right)\\
   \nonumber & \hspace*{1 cm} + P\left((N_{i}^{n}+1) \le c_{K}(1-\epsilon '') n \right)\\\
   \label{eqn: last term eqn 1} & = 0+ P\left((N_{i}^{n}+1) \le c_{K}(1-\epsilon '') n \right)\\
   \label{eqn: last term eqn 2} & \le \exp({-\frac{\epsilon''n}{2}}).
\end{align}
Equality (\ref{eqn: last term eqn 1}) follows from (\ref{eqn: last term equal zero}). From Proposition \ref{prop:lambda star bounded away from zero}, we recognise that $(N_{j'}^{n} - nc_{K})$ is a bounded difference sub-martingale for all $j'$. Hence, inequality (\ref{eqn: last term eqn 2}) follows from the Azuma-Hoeffding inequality for bounded difference sub-martingales. Note that only the last term in (\ref{eqn: exponential bound 1}) is dependent on $L$. By the choice of $\epsilon '$ and for all $n$ under consideration, and from (\ref{eqn: last term eqn 2}), we have shown that it decays exponentially with $n$, and independent of $L$.

It now suffices to show that each of the other terms in (\ref{eqn: exponential bound 1}) decays exponentially with $n$. Let us now look at the first term in (\ref{eqn: exponential bound 1}).
\begin{align}
 \nonumber P & \left( (N_{i}^{n}+1) \left(D\left(\frac{Y_{i}^{n}}{N_{i}^{n}+1} \Vert \frac{Y^{n}-Y_{j}^{n}}{n-N_{j}^{n}}\right)- D(R_{1} \Vert R'_{min})\right) < -\epsilon' n \right)\\
 \nonumber & \le P \left( (N_{i}^{n}+1) \left(D\left(\frac{Y_{i}^{n}}{N_{i}^{n}+1} \Vert \frac{Y^{n}-Y_{j}^{n}}{n-N_{j}^{n}}\right)- D(R_{1} \Vert R'_{min}) \right) < -\epsilon' n, ~N_{j'}^{n} \ge c_{K}(1-\epsilon '')n ~ \forall j' \right)\\
  \label{eqn: first probabilty term 1} & \hspace{1 cm} + \sum_{j'}P\left(N_{j'}^{n} < c_{K}(1-\epsilon '')n \right).
\end{align}
All the terms inside the summation in (\ref{eqn: first probabilty term 1}) have exponential bounds from Proposition \ref{prop:lambda star bounded away from zero} and from Azuma-Hoeffding inequality for bounded difference sub-martingales. The first term in (\ref{eqn: first probabilty term 1}) can be further upper bounded by,
\begin{align}
 \nonumber P &\left((N_{i}^{n}+1)\left(D\left(\frac{Y_{i}^{n}}{N_{i}^{n}+1} \Vert \frac{Y^{n}-Y_{j}^{n}}{n-N_{j}^{n}}\right)- D(R_{1} \Vert R'_{min})\right) < -\epsilon' n, ~ N_{j'}^{n} \ge c_{K}(1 - \epsilon'')n ~ \forall j' \right)\\
 \nonumber & \le P \left((N_{i}^{n}+1)  \left( D\left(\frac{Y_{i}^{n}}{N_{i}^{n}+1} \Vert   \frac{Y^{n}-Y_{j}^{n}}{n-N_{j}^{n}} \right)- D\left(R_{1} \Vert  \frac{Y^{n}-Y_{j}^{n}}{n-N_{j}^{n}} \right)\right) < -\epsilon' n,\right.\\
 \nonumber & \hspace*{4 cm} \left. N_{j'}^{n} \ge c_{K}(1 - \epsilon'')n ~ \forall j', ~ \frac{Y^{n}-Y_{j}^{n}}{n-N_{j}^{n}} \ge R'_{min}, ~ \frac{Y^{n}-Y_{j}^{n}}{n-N_{j}^{n}} \le R_{2} \right)\\
 \label{eqn: exponential bound term1 1}& \hspace*{1 cm} + P \left(\frac{Y^{n}-Y_{j}^{n}}{n-N_{j}^{n}} < R'_{min}, ~ N_{j'}^{n} \ge c_{K}(1 - \epsilon'')n ~ \forall j' \right) + P \left(\frac{Y^{n}-Y_{j}^{n}}{n-N_{j}^{n}} > R_{2}, ~ N_{j'}^{n} \ge c_{K}(1 - \epsilon'')n ~ \forall j' \right).
 \end{align}
 Inequality (\ref{eqn: exponential bound term1 1}) follows by replacing $D(R_{1} \Vert R'_{min})$ by a larger $D\left(R_{1} \Vert  \frac{Y^{n}-Y_{j}^{n}}{n-N_{j}^{n}} \right)$ using the fact that $D(x \Vert y)$ is monotonically increasing in $y$ for $y > x$. Let us now consider the first term in (\ref{eqn: exponential bound term1 1}). Recognise that we have restricted $ \frac{Y^{n}-Y_{j}^{n}}{n-N_{j}^{n}}$ to lie in a compact interval $[R'_{min}, R_{2}]$. Further, since $D(x \Vert y)$ is jointly continuous in $(x,y)$ and since the second argument is restricted to a compact set, we can upper bound the first term in (\ref{eqn: exponential bound term1 1}), for a suitable $\delta_{\epsilon}$, by 
 \begin{align}
 \nonumber P &\left((N_{i}^{n}+1)  \left( D\left(\frac{Y_{i}^{n}}{N_{i}^{n}+1} \Vert   \frac{Y^{n}-Y_{j}^{n}}{n-N_{j}^{n}} \right)- D\left(R_{1} \Vert  \frac{Y^{n}-Y_{j}^{n}}{n-N_{j}^{n}} \right)\right) < -\epsilon' n,\right.\\
 \nonumber & \hspace*{4 cm} \left. N_{j'}^{n} \ge c_{K}(1 - \epsilon'')n ~ \forall j', ~ \frac{Y^{n}-Y_{j}^{n}}{n-N_{j}^{n}} \ge R'_{min}, ~ \frac{Y^{n}-Y_{j}^{n}}{n-N_{j}^{n}} \le R_{2} \right)\\
\label{eqn: exponential bound terms1 2} & \le P \left( \left| \frac{Y_{i}^{n}}{N_{i}^{n}+1} - R_{1} \right| > \delta_{\epsilon},~ N_{j'}^{n} \ge c_{K}(1 - \epsilon'')n ~ \forall j' \right). 
 \end{align}
We recognise that (\ref{eqn: exponential bound terms1 2}) can be expressed as the probability of the deviation of a martingale difference sequence from zero, which we know can be exponentially bounded using the martingale concentration bounds of De la Pena \cite[Theorem 1.2A]{ref:victor1999general}, given in (\ref{eqn: De la Pena martingale concentration result})
 
 Let us define $R''_{min} := R'_{min}+c_{K}\epsilon''(R_{2}-R_{1})$ and $R''_{max} := R'_{max}-c_{K}\epsilon''(R_{2}-R_{1})$. Let $\epsilon''' > 0$ be such that $R'_{min} + 2 \epsilon'''< R''_{min}$ and $ R''_{max}+2\epsilon''' < R'_{max}$. We then recognise that, given the event $\{N_{j'}^{n} \ge c_{K}(1 - \epsilon'')n ~ \forall j'\}$, the event $$\left\{\frac{N_{i}^{n}R_{1} + (n-N_{i}^{n}-N_{j}^{n})R_{2}}{n-N_{j}^{n}} \ge (1-c_{K}(1+\epsilon''))R_{1}+c_{K}(1+\epsilon'')R_{2}  = R''_{min}\right\}$$ is also true. Then, the following statements are true
\begin{align}
\nonumber \left\{\frac{Y^{n}-Y_{j}^{n}}{n-N_{j}^{n}} < R'_{min} \right\} & \subseteq \left\{\frac{Y^{n}-Y_{j}^{n}}{n-N_{j}^{n}} < R''_{min}-\epsilon ''' \right\}\\
 \nonumber &  \subseteq  \left\{\frac{Y^{n}-Y_{j}^{n}}{n-N_{j}^{n}} < \frac{N_{i}^{n}R_{1} + (n-N_{i}^{n}-N_{j}^{n})R_{2}}{n-N_{j}^{n}}-\epsilon ''' \right\}\\
 \label{eqn: Yn-Yjn less than R'min event}&  \subseteq  \left\{ \left| \frac{Y^{n}-Y_{j}^{n}}{n-N_{j}^{n}} - \frac{N_{i}^{n}R_{1} + (n-N_{i}^{n}-N_{j}^{n})R_{2}}{n-N_{j}^{n}} \right| > \epsilon ''' \right\}.
\end{align}
 Similarly, given the event $\{N_{j'}^{n} \ge c_{K}(1 - \epsilon'')n ~ \forall j'\}$, we can show that
 \begin{align}
  \label{eqn: Yn-Yjn greater than R2 event} \left\{\frac{Y^{n}-Y_{j}^{n}}{n-N_{j}^{n}} > R_{2} \right\} & \subseteq \left\{ \left| \frac{Y^{n}-Y_{j}^{n}}{n-N_{j}^{n}} - \frac{N_{i}^{n}R_{1} + (n-N_{i}^{n}-N_{j}^{n})R_{2}}{n-N_{j}^{n}} \right| > \epsilon ''' \right\}.
 \end{align}
From (\ref{eqn: Yn-Yjn less than R'min event}) and (\ref{eqn: Yn-Yjn greater than R2 event}), the second and third term in (\ref{eqn: exponential bound term1 1}) can then be upper bounded by 
\begin{align}
\nonumber P & \left(\frac{Y^{n}-Y_{j}^{n}}{n-N_{j}^{n}} < R'_{min}, ~ N_{j'}^{n} \ge c_{K}(1 - \epsilon'')n ~ \forall j' \right) + P \left(\frac{Y^{n}-Y_{j}^{n}}{n-N_{j}^{n}} > R_{2}, ~ N_{j'}^{n} \ge c_{K}(1 - \epsilon'')n ~ \forall j' \right)\\
\label{eqn: exponential bound terms 2 and 3}& \le 2 P \left( \left| \frac{Y^{n}-Y_{j}^{n}}{n-N_{j}^{n}} - \frac{N_{i}^{n} R_{1} +(n - N_{i}^{n} - N_{j}^{n}) R_{2}}{n-N_{j}^{n}} \right| > \epsilon ''',~ N_{j'}^{n} \ge c_{K}(1 - \epsilon'')n ~ \forall j' \right).
\end{align}
Again, we recognise that (\ref{eqn: exponential bound terms 2 and 3}) can be expressed as the probability of the deviation of a martingale difference sequence from zero, which we know can be exponentially bounded using the martingale concentration bounds of De la Pena \cite[Theorem 1.2A]{ref:victor1999general}, given in (\ref{eqn: De la Pena martingale concentration result}). 
 
Let us now look at the other terms in (\ref{eqn: exponential bound 1}). The second term is identically zero, as the left-hand side is always positive. Arguments similar to those of the first term hold for the third and fourth terms. For the fifth and sixth terms, the left-hand sides converge to a constant, while the right-hand side goes to negative infinity, and thus its straightforward to obtain exponential bounds for these terms. Similarly, for the seventh and eight terms, the left-hand side goes to negative infinity at a logarithmic rate, while the right-hand side goes to negative infinity at a faster linear rate, and again it is straightforward to obtain exponential bounds for these terms. This completes the proof for Lemma \ref{lemma: exponential bound for Zi(n)}.
\end{IEEEproof}

This completes the proof of our main achievability result of Proposition \ref{prop: upper bound on expected stopping time}.
\hfill \IEEEQEDclosed

\bibliographystyle{../../IEEEtran/bibtex/IEEEtran}
{
\bibliography{../../IEEEtran/bibtex/IEEEabrv,../../BIB/ISITbib}
}

\end{document}